\documentclass[11pt]{article}
\usepackage{graphicx}
\usepackage{amssymb}
\usepackage{epstopdf}
\usepackage{color}
\usepackage{amsmath}
\usepackage{mathtools}

\newcommand{\nco}{\newcommand}
\nco{\one}{\ensuremath{\,\,\mathrm{l}\!\!\!1}} 
\nco{\ZZ}{\mathbb{Z}}
\nco{\CC}{\mathbb{C}}
\nco{\RR}{\mathbb{R}}
\nco{\red}{\color{red}}
\nco{\blue}{\color{blue}}
\nco{\magenta}{\color{magenta}}
\nco{\violet}{\color{violet}}
\nco{\redend}{\normalcolor}
\nco{\0}{\underline{0}}

\def\Blue#1{\blue #1\normalcolor}
\definecolor{violet}{rgb}{1,0,1}

\def\ie{{\it i.e.,\/}\ }%
\def\ie{{\rm i.e.,\/}\ }

\nco{\rnc}{\renewcommand}
\rnc{\title}[1]{{\Large\bf\mbox{}\\\medskip#1\bigskip\medskip\\}}
\rnc{\author}[1]{{\large #1\smallskip\\}}
\nco{\address}[1]{{\em #1\medskip\\}}
\def\be{\begin{equation}}\def\ee{\end{equation}}
\def\bea{\begin{eqnarray}}\def\eea{\end{eqnarray}}
\def\bee{\begin{enumerate}}\def\eee{\end{enumerate}}
\def\bei{\begin{itemize}}\def\eei{\end{itemize}}

\def\ommit#1{{}}
\def\su{{\rm su}}
\def\SU{{\rm SU}}
\def\inv#1{\frac{1}{#1}}
\def\mult{{\rm mult}}
\def\tr{\mathbf{t}}

\def\eq=#1{\buildrel #1 \over{=}}

\newtheorem{theorem}{Theorem}

\def\subsubsubsection#1{\medskip{\bf #1}}

 \def\a{\alpha}
 \def\oblade{O-blade}
\def\pictograph{pictograph} 

\begin{document}
\begin{titlepage}
\vspace*{\fill}

\begin{center}
\title{ Conjugation properties of tensor product multiplicities}
\medskip
\author{Robert Coquereaux} 
\address{Centre de Physique Th\'eorique (CPT),\\ 
Aix Marseille Universit\'e, Universit\'e de Toulon, CNRS, CPT, UMR 7332, 13288 Marseille, France}

\medskip
\author{Jean-Bernard Zuber}
\address{
 Sorbonne Universit\'es, UPMC Univ Paris 06, UMR 7589, LPTHE, F-75005, 
Paris, France\\
\& CNRS, UMR 7589, LPTHE, F-75005, Paris, France
 }
\bigskip\medskip

\begin{abstract}
\noindent  {It was recently proven that the total multiplicity in the decomposition into irreducibles of the tensor
product $\lambda\otimes \mu$ of two irreducible representations of a simple Lie algebra is invariant under
conjugation of one of them{;} at a given level, this also applies to the fusion multiplicities of affine algebras{.}
Here, we show that, in the case of SU(3), the lists of multiplicities, in the tensor products 
$\lambda \otimes \mu$ and $\lambda \otimes \overline{\mu}$, are identical up to permutations. 
This latter property does not hold in general for other Lie algebras.
We conjecture that the same property should hold for the fusion product of the affine algebra of su(3) at finite levels, but this is not investigated in the present paper.  }

\end{abstract}
\end{center}
\vspace*{20mm}

\vspace*{\fill}
\end{titlepage}

\section{Introduction}
\subsection{Foreword and Outline}
In a recent paper \cite{CZ1} we have proved the following theorem
{\theorem\label{CZthm}{Let $N_{\lambda\mu}^\nu$ denote the multiplicit{y} 
in the tensor product decomposition
$\lambda\otimes \mu\to \nu$. Then the total multiplicities in $\lambda\otimes \mu$ and 
$\lambda\otimes \bar\mu$ are equal
\be\label{CZ1}\sum_\nu N_{\lambda\mu}^\nu=\sum_{\nu'} N_{\lambda\bar\mu}^{\nu'}\,. \ee}}
This theorem applies to the tensor product category of finite dimensional representations of simple Lie 
groups or algebras, and also to the fusion category of simple affine Lie algebra at finite level $k$.

In the present note {we present a refined version of this conjugation property. }

First let us note that there is another obvious conjugation  identity (of even wider validity) 
\be\label{CZ2}\sum_\nu (N_{\lambda\mu}^\nu)^2=\sum_{\nu'} (N_{\lambda\bar\mu}^{\nu'})^2\,. \ee
Proof. The number of invariants $N_{\lambda\mu \bar\lambda\bar\mu}^{0}$ in 
${\lambda\otimes \mu\otimes\bar\lambda\otimes\bar\mu}$ may be written as
\bea\nonumber N_{\lambda\mu \bar\lambda\bar\mu}^{0}&\eq={\rm {(i)}}&
\sum_{\nu,\nu'} N_{\lambda\mu}^\nu N_{\bar\lambda\bar\mu}^{\nu'}N_{\nu\nu'}^0
\eq={\rm {(ii)}}\sum_{\nu,\nu'} N_{\lambda\mu}^\nu N_{\bar\lambda\bar\mu}^{\nu'}\delta_{\nu'\bar\nu}
\eq={\rm {(iii)}}\sum_\nu N_{\lambda\mu}^\nu N_{\bar\lambda\bar\mu}^{\bar\nu}= \sum_\nu (N_{\lambda\mu}^\nu)^2\\
&\eq={\rm {(iv)}}&N_{\lambda\bar\mu \bar\lambda\mu}^{0}=\sum_\nu N_{\lambda\bar\mu}^\nu N_{\bar\lambda\mu}^{\bar\nu}=\sum_\nu (N_{\lambda\bar\mu}^\nu)^2 \eea
where we have made use of (i) associativity, (ii) $N_{\nu\nu'}^0=\delta_{\nu'\bar\nu}$,  (iii) invariance under conjugation $N_{\bar\lambda\bar\mu}^{\bar\nu}=N_{\lambda\mu}^\nu$, and (iv) commutativity $N_{\lambda\mu \bar\lambda\bar\mu}^{0}=N_{\lambda\bar\mu \bar\lambda\mu}^{0}$ .

In view of the two sum rules (\ref{CZ1}-\ref{CZ2}),   one could think  that the lists of multiplicities
in the tensor (or fusion) products $\lambda\otimes \mu$ and $\lambda\otimes \bar\mu$ might be 
identical (up to permutations). It turns out that this is not true in general, except for the
case of SU(3).

{\theorem\label{CZthm2}{The lists of multiplicities of SU(3)  h.w. appearing in the decomposition into 
irreps of $\lambda\otimes \mu$ and $\lambda\otimes \bar\mu$ are the same (up to permutation).}}

That the theorem fails in general for {other simple Lie algebras}
 may be seen by explicit counterexamples, see sect. \ref{final-comments}.

A first 
proof in sect.\,\ref{wellknownfacts} relies on ``well known" facts on multiplicities in the decomposition of tensor products
of irreps of SU(3), combined with Theorem \ref{CZthm}. Fifty years ago, Mandel'tsveig \cite{Mandeltsveig}, motivated by 
the physics of hadrons and what is called nowadays ``flavor SU(3)", gave a detailed discussion of irreps
of SU(3), including the decomposition rules of tensor products. We shall see in sect. \ref{wellknownfacts} that using either
his results, or Wesslen's more recent  discussion of the same issue \cite{Wesslen}, we can derive  Theorem \ref{CZthm2}
from Theorem \ref{CZthm}.  \\
Another proof (sect.\,\ref{proofusingKT}) is more  detailed and technical, since it gives a description of a mapping from 
the representations $(\nu,\a)$ appearing in $\lambda\otimes \mu$, with $\a$ a multiplicity index,
to those $(\nu',\a')$ appearing in $\lambda\otimes \bar\mu$. We find it convenient to adopt the 
combinatorial picture where the multiplicity $\a$ is regarded as an integral index satisfying a set of inequalities.
Several such pictures have been proposed in the literature (B-Z triangles, K-T honeycombs, hives or
puzzles, \oblade s, etc), some of which we review briefly in sect.\,\ref{combinatorialmodels} for the convenience of the reader. The mapping 
$(\nu,\a) \mapsto (\nu',\a')$ that we propose (a priori it is non unique) turns out to be piecewise linear. 
In our final section  \ref{final-comments}, we comment briefly on what goes wrong in higher rank cases.

\def\p{\lambda_1}
\def\q{\lambda_2}
\def\r{\mu_1} 
\def\s{\mu_2} 
\def\x{\nu_1} 
\def\y{\nu_2}

\subsection{{About intertwiners}}
An intertwiner of the group $G$ is an element of  $(V_1 \otimes V_2 \otimes V_3)^G$, \ie a $G$-invariant vector of this tensor product, where the $V_j$ are representations spaces for $G$.
Such an element can be considered as an equivariant map $V_1 \otimes V_2 \rightarrow V_3^\star$ where the last term denotes the dual of $V_3$ (with the action of $G$ given by the contragredient representation).
If the $V_j$ are irreducible representations, the dimension $d$ of this space of intertwiners is called the multiplicity of $V_3^\star$ in the decomposition of $V_1 \otimes V_2$ into irreducible representations (irreps in short). 
We say that $V_3^\star$ occurs in $V_1 \otimes V_2$ (or that  $V_3^\star$ is a component of $V_1 \otimes V_2$) if $d>0$.
Given the action of $G$ on $V_1 \otimes V_2$, we may consider its centralizer algebra (Schur), or, equivalently, the commutant of $G$ in the space of endomorphisms  $End(V_1 \otimes V_2)$.  
If the multiplicity of $V_3^\star$ is $d$, the centralizer algebra of $G$ in this space of endomorphisms contains a simple block $d \times  d$.

Let $G$ be a semi-simple Lie group. In this framework the numbers $m_{\lambda \mu}^\nu$ giving the multiplicity of $V_\nu$ in the tensor product $V_\lambda \otimes V_\mu$ are called  Littlewood-Richardson coefficients for $G$.  As irreps of $G$ are highest weight representations we can characterize the three vector spaces $V_\lambda$, $V_\mu$, $V_\nu$ by a triple $(\lambda,\mu,\nu)$ of (integral) dominant weights. The notation $\lambda\otimes \mu\to \nu$ or, equivalently, $\lambda\otimes \mu \otimes \nu^\star \to \one$, will refer to the corresponding space of intertwiners.

 The ``3J-operators'' of physicists are intertwiners, thought of  as linear maps from  $V_\lambda \otimes V_\mu  \otimes V_\nu^\star$ to the scalars. Once evaluated on a triple of vectors, such a
 3J operator becomes a 3J symbol (a number), also called a Clebsh-Gordan coefficient. 

 Combinatorial models typically associate  to  a space of intertwiners of dimension $d$, and therefore to a given triple\footnote{The order matters:  despite the trivial symmetry $m_{\lambda \mu}^\nu = m_{\mu \lambda }^\nu$ coming from a representation theory interpretation, the two families of objects  
 associated with the intertwiners spaces describing  $\lambda \otimes \mu \rightarrow \nu$ and  
$\mu \otimes \lambda \rightarrow \nu$ are different.} $(\lambda, \mu, \nu)$, a set of $d$ combinatorial and graphical objects,
that we call generically {\it \pictograph s}\footnote{Webster says : ``{\bf pictograph}.  1 :  an ancient or prehistoric drawing or painting on a rock wall ; 2 :  one of the symbols belonging to a pictorial graphic system; etc.".  Here we are more concerned with 
the pictorial graphic system than with the prehistoric graffiti\dots}. 
Examples are provided by the Knutson-Tao (KT) honeycombs.  
There are other models, whose definition is recalled in section \ref{combinatorialmodels},
 that can be used in the same way as KT-honeycombs, {but the proof of the permutation property that we shall present in section \ref{proofusingKT} will use the latter}.
 For {our} purposes, {we do not} need to establish a precise correspondance between {specific} intertwiners and \pictograph s, 
it is enough to remember that one can associate, with the (ordered) triple of weights labelling a space of intertwiners, a collection of $d$ distinct \pictograph s if this space of intertwiners is of dimension $d$.


\subsection{Notations}
\label{notations}
In the present paper, we shall be mainly dealing with the SL(3) and SU(3) Lie groups and algebras.
Their irreducible finite dimensional representations  are labelled by their dominant integral highest weights $\lambda = \lambda_1 \omega_1 + \lambda_2 \omega_2$,
where $\{\omega_1,\omega_2\}$ are the fundamental weights, and $\lambda$ may be given by its Dynkin labels $(\lambda_1,\lambda_2)$, where $0 \leq \lambda_1, \lambda_2 \in \ZZ$.
Occasionally we shall make use of the three 
weights $\epsilon_i$, $i=1,2,3$, of the $\omega_1$ representation (they are non-independent,  $\sum \epsilon_i=0$)
and of the simple roots $\alpha_1=(\epsilon_1-\epsilon_2)$, $\alpha_2=(\epsilon_2-\epsilon_3)$. \\
In terms of the $\epsilon$'s we have 
\be 
\epsilon_1\cdot \lambda=\inv{3}(2\lambda_1+\lambda_2)\ ;\ \ 
\epsilon_2\cdot \lambda=-\inv{3}(\lambda_1-\lambda_2)\ ;\ \  
{\epsilon_3 \cdot \lambda = -\inv{3}(\lambda_1+2\lambda_2)}\,, 
\ee
and thus $\lambda_1=(\epsilon_1-\epsilon_2) \cdot \lambda$, \ $\lambda_2=(\epsilon_2-\epsilon_3) \cdot \lambda$. 

 {In SU(3) there is  a $\ZZ_3$ grading $\tau$ (``triality'') on irreps, 
stemming from the fact that this discrete group is the center of SU(3); we set  
$\tau({\omega_1})=1, \tau({\omega_2})=2\; \textrm {mod} \; 3$
and more generally 
\be \label{triality} \tau(\lambda):=\lambda_1+2\lambda_2 \; \textrm{mod} \; 3\,. \ee
This triality is conserved (mod 3) in the decomposition of tensor products : $N_{\lambda\mu}^\nu\ne 0$ implies
 \be\label{cons-trial}\tau(\lambda)+\tau(\mu)-\tau(\nu) =0\ \mod 3\,.\ee}

 \subsection{Miscellaneous comments}
  
  We should stress the fact that the property of multiplicities described by theorem 2 is discussed only in the framework of the Lie group SU(3). 
  In particular, we do not discuss the analogous property for the affine algebra $\widehat{su}(3)_k$, although we conjecture that it should hold as well.
  As already mentioned, the higher rank case, for simple Lie groups or for the corresponding affine algebras, is briefly commented in one paragraph of our final section (in those cases theorem 1 holds but theorem 2 does not).

 There are many references in the literature about fusion rules and tensor product multiplicities, see for instance \cite{DFMS} and references therein.  
 We should probably mention at that point that, in the mathematical literature, there is a result about tensor multiplicities known as the PRV conjecture \cite{PRV}.
  This result, which is usually expressed in terms of the geometry of the Weyl group,  gives a sufficient condition, for a dominant weight, to appear as the highest weight of an irreducible representation contained in the tensor product of two given irreps.  This conjecture became a theorem when it was independently proved in \cite{Kumar} and \cite{Mathieu}. It was later generalized in \cite{MPR}.  
 In the framework of affine algebras, a similar result for the fusion product was obtained in \cite{FrenkelZhu}; actually, it was only after the work of \cite{FeingoldFredenhagen} that the result of \cite{FrenkelZhu} could be recognized as a    generalization of the PRV theorem to the affine case.
Our result (theorem 2), although leading to severely contrived conditions on the irreps that may occur in a  tensor product of two given irreducible representations, should not be considered as a generalization of the PRV theorem.
Indeed, although it is, in a sense, stronger than the latter result, it only applies to the Lie group SU(3).

\def\s{\mu_2} 

\section{A proof based on ``well known" facts on SU(3) multiplicities}
\label{wellknownfacts}

\def\k{\ell}
\subsection{Mandeltsveig's results on multiplicities} 
\label{mandel}
Consider now the SU(3) group and a pair of irreps of h.w. $\lambda$ and $\mu$. 
We assume that we have ordered the two given weights $\lambda$ and $\mu$ in such a way that ${\max}(\lambda_1,\lambda_2) \geq {\max}(\mu_1,\mu_2)$.
For every positive (or null) integer $\k$, 
we set $n(\lambda,\mu; \k) = \min(\lambda_1-\k,\mu_2) + \min(\mu_1-\k,\lambda_2)$. 
For every negative integer $\k$, we set 
$n(\lambda,\mu; \k) ={\min}(\lambda_1,\mu_2+\k) + {\min}(\mu_1,\lambda_2+\k)$.
The number $M$ of distinct h.w. (\ie inequivalent irreducible representations)  appearing in the tensor product $\lambda \otimes \mu$ is given by the following formula \cite{Mandeltsveig}
\begin{equation}
\label{distinctmultiplicities}
M=\sum_{\k= -\min(\lambda_2, \mu_2) }^{\min(\lambda_1, \mu_1)} (n(\lambda_1,\lambda_2,\mu_1, \mu_2;\k)+1)\,.
\end{equation}

Call $\rm{mult}({\lambda, \mu; \nu})$ the multiplicity of $\nu$ in $\lambda \otimes \mu$  
and $\rm{mult}_{max}(\lambda, \mu)$ its maximal value. Then 
\begin{equation}
\label{maximalmultiplicity}
\rm{mult}_{max}(\lambda, \mu) = \min({\lambda_1, \lambda_2, \mu_1, \mu_2)}  + 1\,.
\end{equation}  
It may happen that $\rm{mult}({\lambda, \mu; \nu}) = \rm{mult}({\lambda^\prime, \mu^\prime; \nu^\prime})$ for $\nu \neq \nu^\prime$.
Let us therefore call $\sigma(\lambda, \mu;s) = \# \{\nu \; \textrm {such that} \;  \textrm{mult}(\lambda, \mu; \nu) = s \}$ the number of distinct irreps that appear with multiplicity $s$ in the tensor product $\lambda \otimes \mu$.  From results in \cite{Mandeltsveig},  one can easily deduce the following:
\be\label{termswithgivenmultiplicity2}
\sigma(\lambda, \mu;s)=\begin{cases} \min(\delta\lambda_1, \delta\mu_1 {+} \delta\mu_2)+ \min( \delta\lambda_2, \delta\mu_1 {+}\delta\mu_2) + \delta\mu_1 + \delta\mu_2 &{\rm if}\ \  s< \rm{mult}_{max}(\lambda, \mu) \\
M-\sum_{s=1}^{ \rm{mult}_{max}(\lambda, \mu)-1} \sigma(\lambda, \mu;s) & {\rm if}\ \  s=\rm{mult}_{max}(\lambda, \mu)
\end{cases}
\ee
where for $n$ an arbitrary number,  $\delta n := n - (s-1)$. Note that 
each $\delta n$ in that expression is strictly positive, and so is the sum, showing that, as $\nu$ varies, the multiplicity takes all consecutive values $1, 2, \ldots, \rm{mult}_{max}(\lambda, \mu)$.

{Example. Consider $\lambda=(21,6)$ and $\mu=(17,16)$. 
One finds from (\ref{maximalmultiplicity}) that their tensor product contains irreps with multiplicities up to $7$.
The numbers of inequivalent irreps with multiplicities $\{1,2,\ldots,6\}$, as given by (\ref{termswithgivenmultiplicity2}), are respectively $\{60,56,52,48,44,40\}$, whose sum is $300$.
 The total number of distinct irreps, given by (\ref{distinctmultiplicities}), is $411$. The number of inequivalent irreps with multiplicities equal to $7$, again given by (\ref{termswithgivenmultiplicity2}), is therefore $411-300=111$.} 
 {See the plot below in Fig. \ref{multip}.} 
 
 
 \subsection{A second proof of the permutation property.}

On the one hand, the formula  (\ref{termswithgivenmultiplicity2}) giving the number of terms with prescribed multiplicity, 
strictly smaller than the maximal one,
is obviously symmetric if we interchange $\mu$ and $\overline \mu$ (\ie $\mu_1$ and $\mu_2$). 
The maximal multiplicity (and hence the list of possible multiplicities), given by equation (\ref{maximalmultiplicity}) stays also the same for the same reason. 
On the other hand, the total multiplicity is also the same, because of theorem \ref{CZthm}, for $\lambda \otimes \mu$ and $\lambda \otimes \overline \mu$.
We conclude that the number of distinct h.w. with maximal multiplicity is also the same in both cases,
or equivalently that the $M$ in (\ref{distinctmultiplicities}) and (\ref{termswithgivenmultiplicity2})  is the same 
{for $\lambda \otimes \mu$ and $\lambda \otimes \overline \mu$}.
From these two results we deduce that, for SU(3), the lists of multiplicities remain the same under the 
``pseudo-conjugation'' $\lambda \otimes \mu \rightarrow \lambda \otimes \overline{\mu}$.


\subsection{Example}

We decompose the tensor product ${(9,5)} \otimes {(2,6)}$ into irreps and compare the result with the 
 {tensor product}  ${(9,5)} \otimes {(6,2)}$. 
 Equations (\ref{distinctmultiplicities}, \ref{maximalmultiplicity}, \ref{termswithgivenmultiplicity2}) imply that, in both cases,  the number of inequivalent irreps occurring in the tensor product(s) is  $M=51$, that the maximal multiplicity is $3$, and that the number of terms with multiplicities $1,2$ is respectively given by $21, 16$. The number of h.w. occurring with maximal multiplicity (namely $3$) is $51 - (21 + 16 ) = 14$. Notice that applying equation (\ref{termswithgivenmultiplicity2}), first line, to the maximal multiplicity would give instead the erroneous result $11$. The total multiplicity is $1 \times 21 + 2 \times 16 + 3 \times 14 = 95$. It is interesting to perform the actual decomposition of the tensor products. Gathering terms by increasing {multiplicities}, one finds:

{\tiny
 $(9,5) \otimes (6,2)= \\
\{\{(1, 11), (2, 9), (2, 12), (3, 7), (3, 13), (4, 5), (5, 3), (5, 
   12), (6, 1), (7, 11), (8, 0), (9, 10), (11, 0), (11, 9), (13, 
   8), (14, 0), (15, 1), (15, 7), (16, 2), (16, 5), {(17,3)}\},
   \\
   \{(3, 10), (4, 
   8), (4, 11), (5, 6), (6, 4), (6, 10), (7, 2), (8, 9), (9, 1), (10, 
   8), (12, 1), (12, 7), (13, 2), (14, 3), (14, 6), {(15,4)}\}, 
   \\
   \{(5, 9), (6, 
   7), (7, 5), (7, 8), (8, 3), (8, 6), (9, 4), (9, 7), (10, 2), (10, 
   5), (11, 3), (11, 6), (12, 4), {(13,5)}\}\}\,.
 $
} 
 
 {\tiny
$(9,5) \otimes (2,6)= \\
\{\{(1, 7), (2, 5), (2, 8), (3, 3), (3, 9), (4, 10), (5, 2), (5, 
   11), (6, 12), (7, 1), (7, 13), (9, 0), (9, 12), (11, 11), (12, 
   0), (12, 9), (13, 7), (14, 5), (15, 0), (15, 3), {(16,1)}\},
   \\
    \{(3, 6), (4, 
   4), (4, 7), (5, 8), (6, 3), (6, 9), (7, 10), (8, 2), (8, 11), (10, 
   1), (10, 10), (11, 8), (12, 6), (13, 1), (13, 4), {(14,2)}\},
   \\
    \{(5, 5), (6, 
   6), (7, 4), (7, 7), (8, 5), (8, 8), (9, 3), (9, 6), (9, 9), (10, 
   4), (10, 7), (11, 2), (11, 5), {(12,3)}\}\}\,.
$
}

\def\u{\lambda}\def\v{\mu}\def\w{\nu}
We display below in
Fig.  \ref{tensorpolygon9562} (a) the ``tensor polygons" of $\u \otimes \v$ {and} $\overline{\u} \otimes  \overline{\v}$,   to be compared with the tensor polygons of  {$\u \otimes \v$} {and} {${\u} \otimes  \overline{\v}$} given in Fig. \ref{tensorpolygon9562} (b).

\begin{figure}[htbp]
\begin{minipage}{16pc}
\centering
\includegraphics[width=15pc]{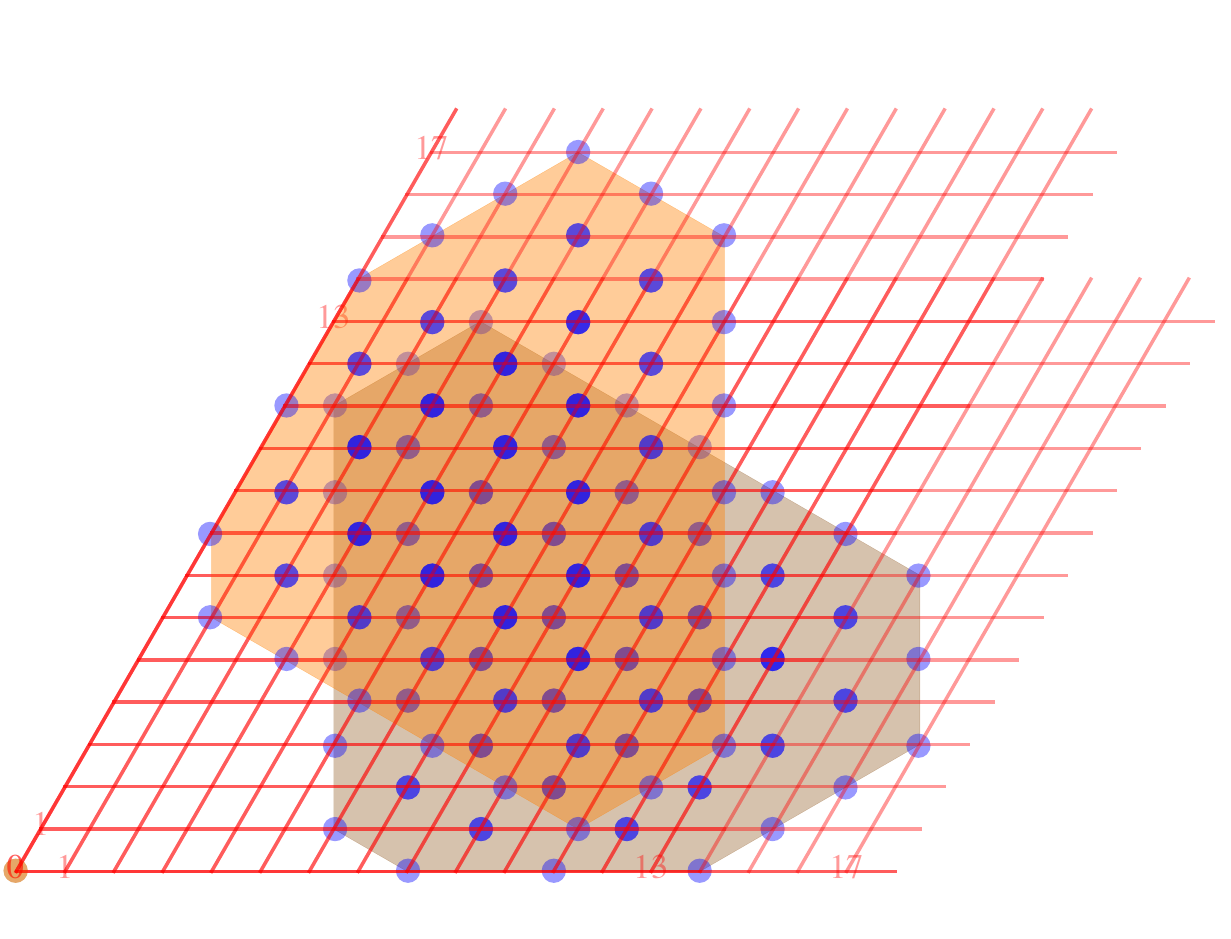}\\
(a)  $\lambda\otimes \mu$ and $\overline{\lambda} \otimes \overline{\mu}$.
\end{minipage}\hspace{2pc}
\hspace{4pc}
\begin{minipage}{16pc}
\centering
\includegraphics[width=15pc]{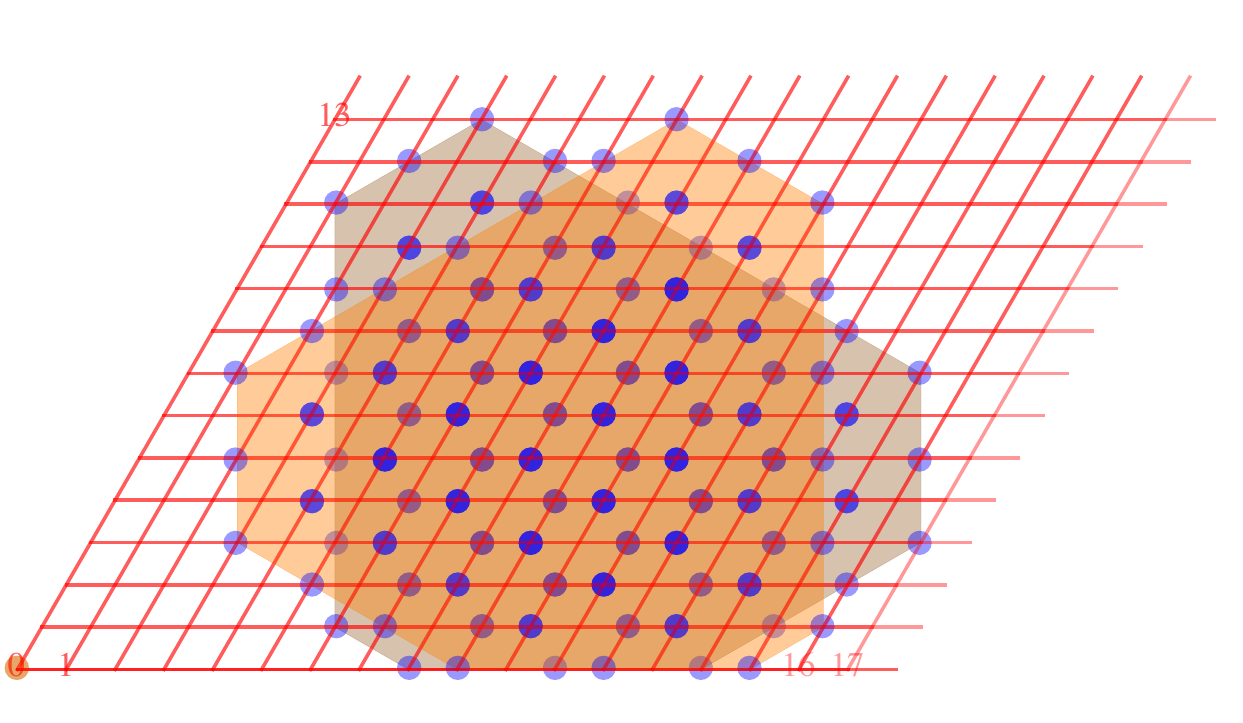}\\
(b)  $\lambda\otimes \mu$ and ${\lambda} \otimes \overline{\mu}$
\end{minipage} 
\caption{\label{tensorpolygon9562}Tensor polygons for $\lambda = (9,5)$ and $\mu=(6,2)$}
\end{figure}

\begin{figure}[tbp]
\centering{\includegraphics[width=20pc]{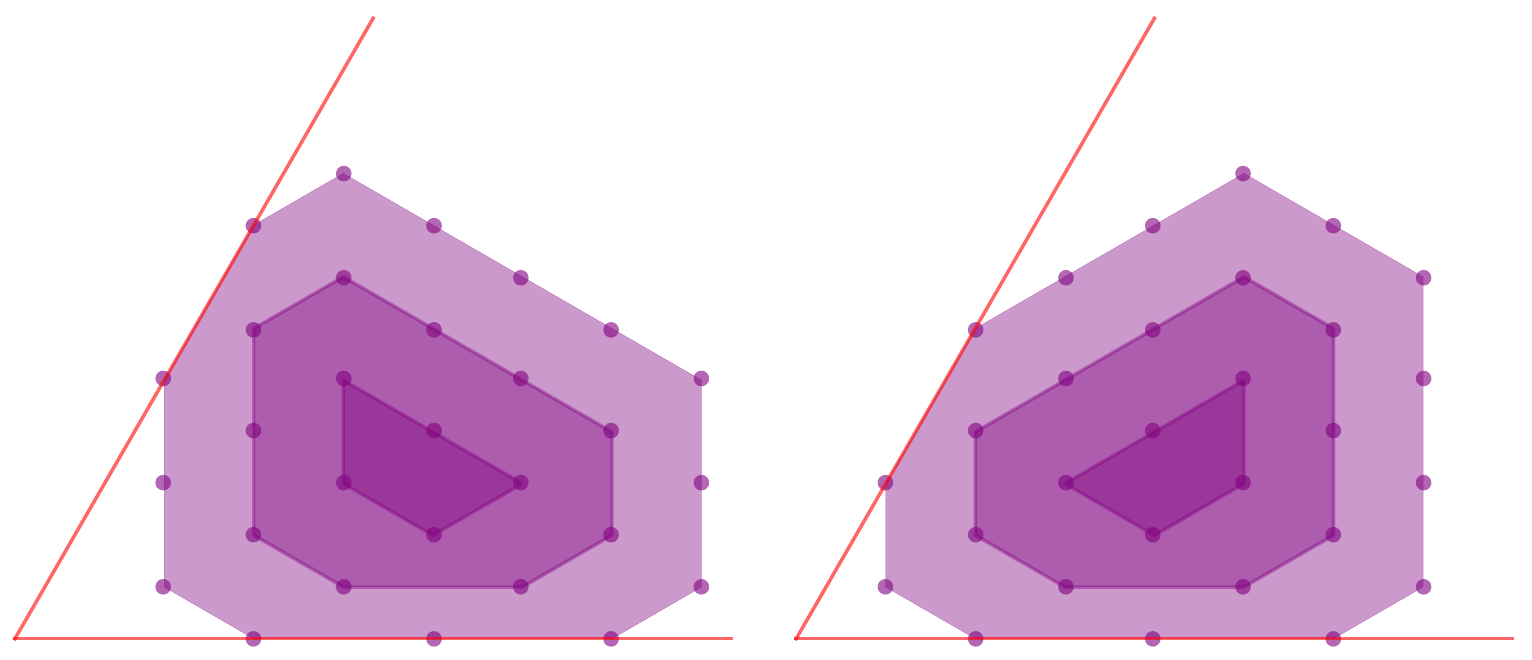}}
\caption{\label{tenspolyu53v42uvanduvbar} Tensor polygons $\u\otimes \v$ and $\u \otimes \overline{\v}$ for $\u=(5,3 )$ and  $\v=(4,2)$}
\end{figure}

In Fig.   \ref{tensorpolygon9562} (a)  as expected,  the two polygons
 are symmetric with respect to the diagonal of the first quadrant. In Fig.  \ref{tensorpolygon9562} (b), the second polygon can be obtained from the first by a {reflection}.
 This last property is not a generic feature as it can be seen for instance from Fig.  \ref{tenspolyu53v42uvanduvbar} that displays the tensor polygons $\u\otimes \v$ and $\u \otimes \overline{\v}$, for $\u=(5,3 )$ and  $\v=(4,2)$: indeed, the sides of the outer shells (convex hulls) do not contain the same number of integer points. A detailed analysis of this last example shows that using a reflection to map the first figure to the next would send one of its vertices outside of the main chamber, so that such a reflection has to be amended.  As we shall see in section \ref{piecewiselinearmap},  the transformation mapping $\u\otimes \v$ to $\u \otimes \overline \v$, is, in the general case, only piecewise linear.

\begin{figure}[htbp]
\begin{minipage}{16pc}
\centering
\includegraphics[width=16pc, angle=145]{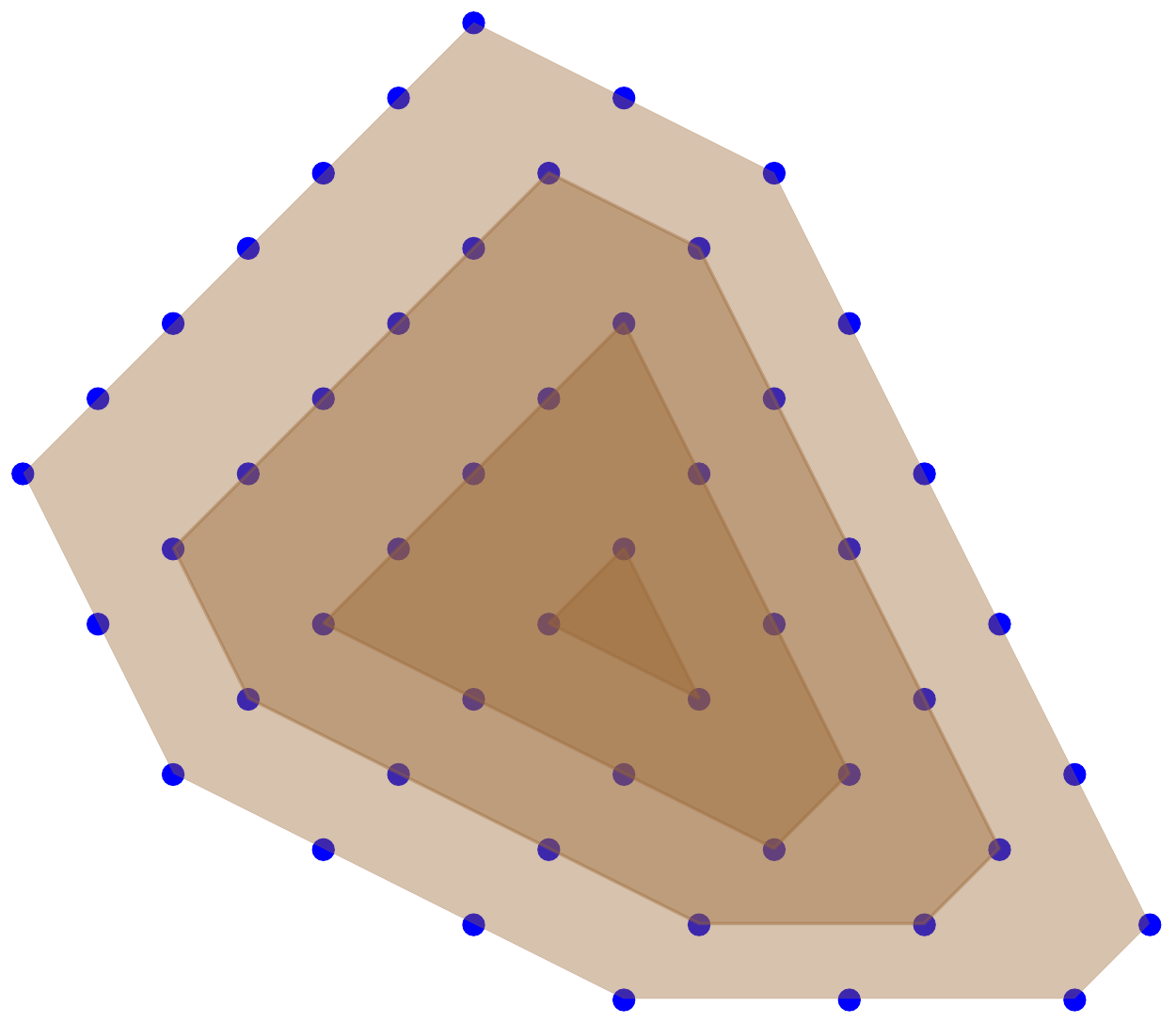}
\end{minipage}
\hspace{3pc}
 \begin{minipage}{16pc}
 \centering
\includegraphics[width=20pc, height = 20pc]{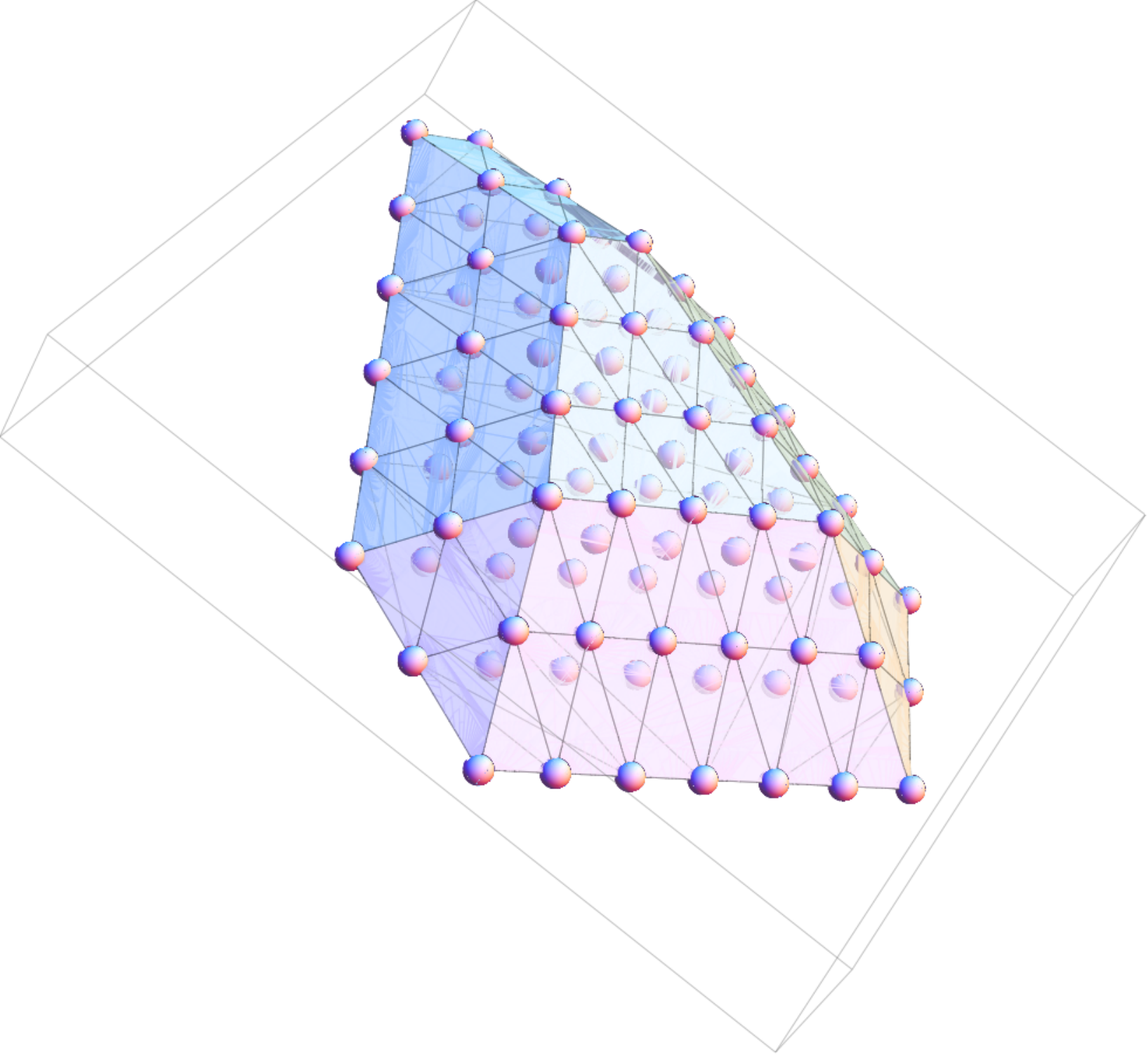}\\
\end{minipage}
\caption{\label{protrudedpolyhedron} The tensor polygon $(9,5) \otimes (6,2)$ of Fig. \ref{tensorpolygon9562} and its hive protruded version. 
}
\end{figure}

Figure \ref{protrudedpolyhedron} gives a protruded 3d version of the previously displayed $(9,5) \otimes (6,2)$ tensor polygon 
(the two {dark} heptagons in   Fig. \ref{tensorpolygon9562} (a) 
and  \ref{tensorpolygon9562} (b))  where the ``degeneracy" 
 of irreps stemming from the non-trivial multiplicities $(1,2,3)$ is lifted by the introduction of a hive parameter 
 discussed below in section \ref{honeycomb-hive}.

\section{{A proof based on KT-honeycombs and hives}}
\label{proofusingKT}
The determination of tensor products multiplicities, aka Littlewood--Richardson coefficients,
is a fascinating subject, with many far reaching implications in various fields of
mathematics and physics. As such it has attracted a lot of attention and has been --and still is--  the object of much ingenuity in 
the development of combinatorial and graphical tools. To quote a few, skew Young tableaux \cite{skYoung},
Berenstein--Zelevinsky triangles \cite{BZ}, honeycombs or their dual hives \cite{KT1}, puzzles \cite{KT2},
``\oblade s" \cite{AOblades}, Littelmann paths \cite{Littelmann}, etc. Each of those has its own merits and may be best suited for a specific problem. 
The discussion that follows can be carried out with various combinatorial models {(various pictographs)}, but for definiteness, 
in this section, we present a proof of our result about the conjugation properties of tensor products  by using the honeycombs, or hives, of Knutson and Tao. 
 We shall indicate, in section \ref{combinatorialmodels}, how the discussion should be modified if we were using other combinatorial models. 

\subsection{Knutson--Tao's honeycombs for  GL(3)}

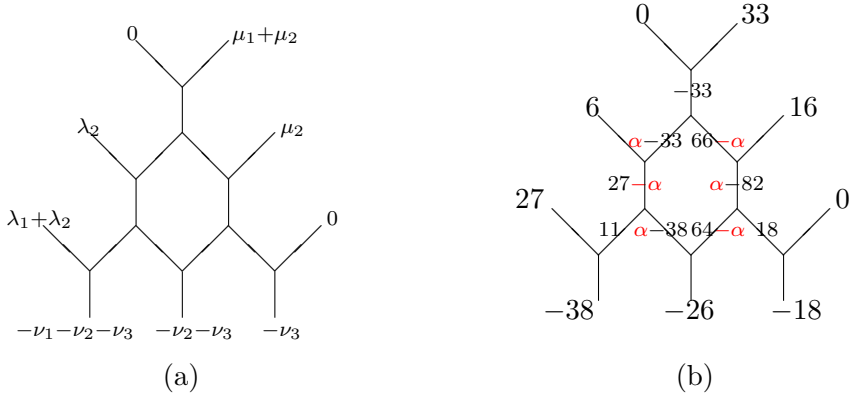
\begin{figure}[ht]
 \setlength{\unitlength}{0.7pt}
 \begin{center}
\begin{picture}(40,80)
                           \put (70,45){$0$}             \put(100,20){\line(1,1){25}}\put(100,20){\line(-1,1){25}} \put (127,45){$33$}
                                                                     \put(100,20){\line(0,-1){25}}    \put(90,5){$\scriptstyle -33$}
                                        \put(66,-22){$\scriptstyle  {\red \a}-33 $}    
                                           \put(100,-22){$\scriptstyle  66{\red {- \a}} $}                           
 \put (43,-5){$6$}  \put(75,-30){\line(-1,1){25}}\put(100,-5){\line(-1,-1){25}}\put(100,-5){\line(1,-1){25}}\put(125,-30){\line(1,1){25}}
 			\put(55,-45){$\scriptstyle   27\red{- \a}$}\put(110,-45){$\scriptstyle  {\red \a}-82$}
                                                       \put(75,-30){\line(0,-1){25}} \put(125,-30){\line(0,-1){25}} \put (153,-5){$16$}
 \put (5,-55){$27$}   \put(25,-55){\line(1,-1){25}}    \put(50,-70){$\scriptstyle 11$}
  \put(69,-70){$\scriptstyle {\red \a}-38$}
  \put(100,-70){$ \scriptstyle 64{\red - \a}$}
   \put(135,-70){$ \scriptstyle 18 $} 
  \put(75,-55){\line(-1,-1){25}}\put(75,-55){\line(1,-1){25}}\put(125,-55){\line(-1,-1){25}}\put(125,-55){\line(1,-1){25}}  \put(175,-55){\line(-1,-1){25}} \put (178,-55){$0$}
                 \put(50,-80){\line(0,-1){25}} \put(100,-80){\line(0,-1){25}}\put(150,-80){\line(0,-1){25}} 
                  \put(20,-115){$-38$}         \put(85,-115){$-26$}  \put(143,-115){$-18$}       
   \end{picture}
\end{center}
\setlength{\unitlength}{0.7pt}
\vskip-22mm
\hskip 10mm
\begin{picture}(-400,80)
                           \put (70,45){$\scriptstyle 0$}             \put(100,20){\line(1,1){25}}\put(100,20){\line(-1,1){25}} \put (127,45){$\scriptstyle\mu_1+\mu_2$}
                                                                     \put(100,20){\line(0,-1){25}}   
                                                                      \put (43,-5){$\scriptstyle\lambda_2$}  \put(75,-30){\line(-1,1){25}}\put(100,-5){\line(-1,-1){25}}\put(100,-5){\line(1,-1){25}}\put(125,-30){\line(1,1){25}}
 			                                                       \put(75,-30){\line(0,-1){25}} \put(125,-30){\line(0,-1){25}} \put (153,-5){$\scriptstyle\mu_2$}
 \put (5,-55){$\scriptstyle \lambda_1+\lambda_2$}   \put(25,-55){\line(1,-1){25}}   
   \put(75,-55){\line(-1,-1){25}}\put(75,-55){\line(1,-1){25}}\put(125,-55){\line(-1,-1){25}}\put(125,-55){\line(1,-1){25}}  \put(175,-55){\line(-1,-1){25}} \put (178,-55){$\scriptstyle 0$}
                 \put(50,-80){\line(0,-1){25}} \put(100,-80){\line(0,-1){25}}\put(150,-80){\line(0,-1){25}} 
                  \put(10,-115){$\scriptstyle  -\nu_1-\nu_2-\nu_3$}         \put(85,-115){$\scriptstyle -\nu_2-\nu_3$}  \put(143,-115){$\scriptstyle -\nu_3$}       
              \end{picture}
\vskip3cm
\hskip32mm (a) \hskip62mm (b)
\caption{KT honeycombs for SU(3): (a): the general case, and (b) one relevant for our example $(21,6)\otimes(17,16)\to(12,8)$}
\label{KThoney}
\end{figure}

We consider the tensor product of SU(3)  representations $(\lambda_1,\lambda_2)\otimes(\mu_1,\mu_2)$,
 the multiplicity of the h.w. $\nu=(\nu_1,\nu_2)$ in its decomposition, 
 and the intertwiners  $\lambda\otimes \mu \otimes \bar {\nu} \to 1$, 
where $\bar {\nu} = (\nu_2,\nu_1)$  denotes the representation conjugate to $\nu$. In this section we use the notation $\nu^*$ to denote the corresponding GL(3) partition, but there should be no confusion (and a possible confusion with the notation $\nu^\star$ used in a previous section to denote the contragredient representation would be harmless).
 As explained above in sect.\,\ref{notations}, the notation  $(\lambda_1, \lambda_2)$ refers to the components of the {SU}(3) weight $\lambda$ in the basis of fundamental weights (Dynkin labels).  
We recast this in the GL(3) language  used by Knutson and Tao {\cite{KT1}} (but with slightly different notations).
Then  GL(3) h.w. are denoted by braces\footnote{a partition (Young) of the integer $\lambda_1+2 \lambda_2+3 \lambda_3$, for instance}, $\lambda=\{\lambda_1+\lambda_2+\lambda_3, \lambda_2+\lambda_3,\lambda_3\},
\ \mu=\{\mu_1+\mu_2+\mu_3,\mu_2+\mu_3,\mu_3\}$  
and ${\nu^*} =\{-\nu_3,-\nu_2-\nu_3,-\nu_1-\nu_2-\nu_3\}$, but in what follows we will be mainly interested in the case
$\lambda_3=\mu_3=0$, for application to SU(3) highest weights.
A KT honeycomb is a hexagonal pattern describing an intertwiner 
$\lambda\otimes \mu \otimes \bar {\nu} \to 1$, see Fig. \ref{KThoney}.
Its trivalent vertices are  integral points $(\xi,\eta,\zeta)$ in the plane $\xi+\eta+\zeta=0$ in $\mathbb{R}^3$, 
connected by segments in one of the three directions 
SE $\searrow=(0,-1,1)$, SW $\swarrow=(1,0,-1)$ or N $\uparrow=(-1,1,0)$. 
Thus one of the coordinates $\xi,\eta, \zeta$ of the vertices does not vary along each of these
segments, for instance $\xi$ along the SE direction, and we label each internal edge by that non-varying coordinate.
Therefore at each vertex, the sum of  labels of the three incident edges vanishes.
The values on the boundary edges of a SU(3) honeycomb are fixed
 to be (in clockwise order)  
  $(\lambda_1+\lambda_2, \lambda_2,0, \mu_1+\mu_2,\mu_2,0,
  -\nu_3,-\nu_2-\nu_3,-\nu_1-\nu_2-\nu_3)$, see Fig. \ref{KThoney}\,(a). 
 As a consequence of the vanishing rule at each vertex and of these boundary conditions, 
we have 
\be\lambda_1+2\lambda_2+3\lambda_3+\mu_1+2\mu_2+3\mu_3-(\nu_1+2\nu_2+3\nu_3)=0\,,\ee
which expresses the conservation 
of a ``U(1) charge", $\tilde\tau(\lambda)+\tilde\tau(\mu)=\tilde\tau(\nu)$, with
  $\tilde\tau(\kappa):= \kappa_1+2\kappa_2+3\kappa_3$ for a h.w. $\kappa=\{\kappa_1, \kappa_2,\kappa_3\}$,
  extending to GL(3) the  SU(3) triality $\tau$ introduced above in (\ref{triality}).
Thus when we specialize to SU(3) weights,  $\lambda_3=\mu_3=0$, $\nu_3$ has to be fixed to a {non-zero} (in general)
value
\bea\label{zz}  \nu_3&=&\inv{3}(\lambda_1+2\lambda_2+\mu_1+2\mu_2-\nu_1-2\nu_2)\\ 
\nonumber
&=&\inv{3}(\tau(\lambda)+\tau(\mu)-\tau(\nu)) \,,\eea
an integer by virtue of the conservation of triality  (\ref{cons-trial}). 
  
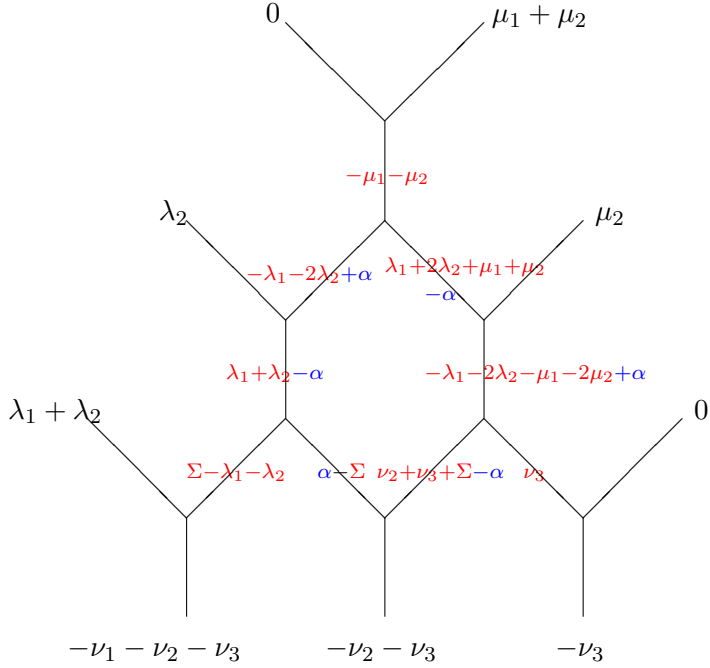
\begin{figure}
\setlength{\unitlength}{1.5pt}
{\begin{picture}(-400,60)
                           \put (120,45){$0$}             \put(150,20){\line(1,1){25}}\put(150,20){\line(-1,1){25}} \put (177,45){$\mu_1+\mu_2$}
                                                                     \put(150,20){\line(0,-1){25}}    \put(140,5){$\red{\scriptstyle -\mu_1-\mu_2}$}
                                        \put(115,-20){$\red{\scriptstyle  -\lambda_1-2\lambda_2 }  \Blue{\scriptstyle + \a} $}    
                                          \put(150,-18){$\red{\scriptstyle \lambda_1+2\lambda_2+\mu_1+\mu_2}$}    \put(160,-25){$\red{\scriptstyle  }{\scriptstyle \Blue{- \a}} $}                           
 \put (93,-5){$\lambda_2$}  \put(125,-30){\line(-1,1){25}}\put(150,-5){\line(-1,-1){25}}\put(150,-5){\line(1,-1){25}}\put(175,-30){\line(1,1){25}}
 			\put(110,-45){$\red{\scriptstyle  \lambda_1+\lambda_2 } {\scriptstyle \Blue{- \a}}$}\put(160,-45){$\red{\scriptstyle    -\lambda_1-2\lambda_2 -\mu_1-2\mu_2} {\scriptstyle\Blue{+\a}}$}
                                                       \put(125,-30){\line(0,-1){25}} \put(175,-30){\line(0,-1){25}} \put (203,-5){$\mu_2$}
 \put (55,-55){$\lambda_1+\lambda_2$}   \put(75,-55){\line(1,-1){25}}    \put(100,-70){$\red{\scriptstyle \Sigma -\lambda_1-\lambda_2}$}
  \put(133,-70){$\scriptstyle \Blue{ \a}-\red{\Sigma}$}\put(148,-70){$\red{\scriptstyle \nu_2+\nu_3+\Sigma}\Blue{\scriptstyle - \a}$} \put(185,-70){$\red{ \scriptstyle \nu_3} $} 
  \put(125,-55){\line(-1,-1){25}}\put(125,-55){\line(1,-1){25}}\put(175,-55){\line(-1,-1){25}}\put(175,-55){\line(1,-1){25}}  \put(225,-55){\line(-1,-1){25}} \put (228,-55){$0$}
                 \put(100,-80){\line(0,-1){25}} \put(150,-80){\line(0,-1){25}}\put(200,-80){\line(0,-1){25}} 
                  \put(70,-115){$ -\nu_1-\nu_2-\nu_3$}         \put(135,-115){$-\nu_2-\nu_3$}  \put(193,-115){$-\nu_3$}       
 \end{picture}    \vskip7cm            
}
\caption{ KT honeycomb for SU(3): the inner edges}
\label{KThoneycombSL3}
\end{figure}

We then compute the coordinates of  vertices, \ie the labels of edges, of such a KT honeycomb. 
There is only one free parameter $\a$ in the central hexagon, see Fig. \ref{KThoney}\,(b) for an example,
and Fig. \ref{KThoneycombSL3} for the generic case. Note that an SU(3) honeycomb depends {\it in fine} on 7 
parameters, the $3\times 2$ independent components of $\lambda,\mu$ and $\nu$ and the $\a$ parameter.

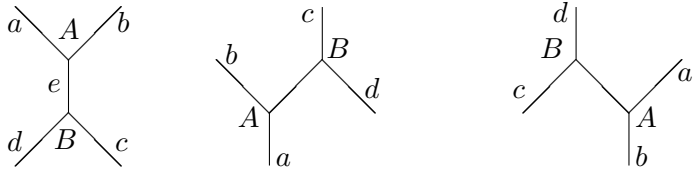
\begin{figure}[ht]
 \vglue-25mm
 \setlength{\unitlength}{0.8pt}
\mbox{
\begin{picture}(0,160)
\put (91,33){$a$}  \put(120,20){\line(1,1){25}}\put(120,20){\line(-1,1){25}} \put (143,33){$b$}\put(115,30){$A$}
 \put(120,20){\line(0,-1){25}}    \put(110,5){$e$}\put(113,-22){$B$}
  \put(120,-5){\line(-1,-1){25}}\put(120,-5){\line(1,-1){25}}
 \put(91,-23){$d$} \put(142,-23){$c$}
   \put(240,45){\line(0,-1){25}} 
  \put(240,20){\line(1,-1){25}}  \put(240,20){\line(-1,-1){25}} \put(215,-5){\line(-1,1){25}} \put(215,-5){\line(0,-1){25}} \put (194,18){$b$}
  \put (260,2){$d$}  \put(200,-12){$A$} \put (218,-30){$a$}  \put (230,38){$c$}  \put(242,21){$B$}
  \put (349,38){$d$}  \put(360,45){\line(0,-1){25}} \put(360,20){\line(-1,-1){25}} \put(360,20){\line(1,-1){25}} \put(385,-5){\line(1,1){25}} \put(385,-5){\line(0,-1){25}} \put (388,-30){$b$} \put (330,2){$c$}  \put (408,10){$a$}
 \put(388,-12){$A$}  \put(343,21){$B$}
           \end{picture}}
\vskip15mm 
\caption{The three  edge patterns}\label{KThoney0}
\end{figure}

There is, however, a further positivity constraint for each of the (nine for SU(3)) internal edges. 
Following Knutson and Tao, we demand that 
edges $\overrightarrow{BA}$ of Fig. \ref{KThoney0} 
be always in one of  the three orientations N, SE or SW (rather than their opposites). For instance in the first edge pattern 
of Fig. \ref{KThoney0}, we have $a+b+e=c+d+e=0$, the points $A$ and $B$ have coordinates $(a,b,-a-b) $ and $(c,d,-c-d)$ respectively, 
and therefore one finds the vector $\overrightarrow{BA}=(a-c,b-d,0)= (b-d) \times \uparrow$. 
We thus impose that  $b\ge d$. 
Likewise,  in the two other patterns of Fig. \ref{KThoney0}, 
$\overrightarrow{BA}= (b-d) \times \swarrow$, resp. $ (b-d) \times \searrow$, 
 and the same condition  $b\ge d$  holds.

For instance, in our favorite example, $\lambda_1=21,\lambda_2=6,\mu_1=17,\mu_2=16$, 
we know  (from the discussion at the end of section \ref{mandel}) that multiplicities range from $1$ to $7$, and that there are $44$ possible $\nu$ for which the multiplicity $\lambda \otimes \mu \rightarrow \nu$  is $5$.
This is the case of $\ \nu_1=12, \nu_2=8$, hence $\nu_3=18$, for which 
 the parameter $\a$ may take 5 values, 
{from 60 to 64},  as  it will be clear in the next section, see Fig. \ref{KThoney}\,(b).

\bigskip

\subsection{Wesslen's inequalities (SU(3)) }
\label{wesslen}

In this subsection we write more explicitly in terms of external variables,
\ie  of components of the weights, the inequalities  that honeycombs must satisfy. In so doing, we make contact with former work
by Wesslen \cite{Wesslen}.
The afore-mentioned positivity constraint is expressed in the case of SU(3) by the 9 inequalities
\bea \nonumber \Sigma-\lambda_1-\lambda_2 \le \a-\lambda_1-2\lambda_2  \le \mu_1+\mu_2 &\quad \quad&\Sigma+ \nu_2+\nu_3-\a\le \mu_2 \\
\label{ineqa}
\Sigma+ \nu_3\le \a \le \Sigma+\lambda_1+\lambda_2 &\quad  \quad&  \lambda_1+2\lambda_2+\mu_1+\mu_2  -\a \le \lambda_2 \\ \nonumber
\nu_3\le  \lambda_1+2\lambda_2+\mu_1+2\mu_2  -\a \le\mu_1+\mu_2 & \quad \quad &  \nu_2+\nu_3\le \a-\lambda_1-\lambda_2\,
\eea
with the notation $\Sigma:=\nu_1+\nu_2+\nu_3$. 
In terms of the triality $\tau$ defined above, 
we see that $\Sigma= \inv{3}(\tau(\lambda) +\tau(\mu)+\tau(\bar\nu))$ whereas (\ref{zz}) reads 
$\nu_3=\inv{3}(\tau(\lambda) +\tau(\mu)-\tau(\nu))=\Sigma-\nu_1-\nu_2$. Both $\Sigma$ and $\nu_3$ are  integers thanks to 
(\ref{cons-trial}).

\noindent Rephrasing these inequalities as 6 lower and 3 upper bounds on $\a$, $a_i\le \a\le b_j$, $i=1,\cdots, 6$ and $1\le j\le 3$, 
we find $3\times 6$ consistency relations $a_i\le b_j$,
namely 
\be \label{systineq} \!\!\!\!\!\!\!\!
\left(
\begin{array}{ccc}
 2 \nu_1+\nu_2\leq 2 \lambda_1+\lambda_2+2 \mu_1+\mu_2 & \lambda_1\geq 0 & \lambda_2+\nu_1\leq \lambda_1+\mu_1+2 \mu_2+\nu_2 \\
 \nu_1+2 \nu_2\leq \lambda_1+2 \lambda_2+\mu_1+2 \mu_2 & \mu_1+\nu_2\leq 2 \lambda_1+\lambda_2+\mu_2+\nu_1 & \mu_2\geq 0 \\
 \mu_2+\nu_1\leq \lambda_1+2 \lambda_2+\mu_1+\nu_2 & \mu_1+2 \mu_2\leq 2 \lambda_1+\lambda_2+\nu_1+2 \nu_2 & \nu_2\geq 0 \\
 \lambda_2\geq 0 & 2 \mu_1+\mu_2\leq \lambda_1+2 \lambda_2+2 \nu_1+\nu_2 & \lambda_1+\mu_1\leq \lambda_2+\mu_2+\nu_1+2 \nu_2 \\
 \mu_1\geq 0 & \lambda_2+\mu_2\leq \lambda_1+\mu_1+2 \nu_1+\nu_2 & \lambda_1+2 \lambda_2\leq 2 \mu_1+\mu_2+\nu_1+2 \nu_2 \\
 \lambda_1+\nu_2\leq \lambda_2+2 \mu_1+\mu_2+\nu_1 & \nu_1\geq 0 & 2 \lambda_1+\lambda_2\leq \mu_1+2 \mu_2+2 \nu_1+\nu_2 \\
\end{array}
\right)\,.\ee   
The inequalities $\lambda_1,\lambda_2,\mu_1,\mu_2\ge 0$ are satisfied by definition. 
We are left with the following inequalities on $\nu_1$ and 
$\nu_2$
\bea  \nonumber   \nu_1\ge 0&\qquad & \nu_2\ge 0\\ \label{8ineq} 
 \scriptstyle \max(2\mu_1+\mu_2-\lambda_1-2\lambda_2, 2\lambda_1+\lambda_2-\mu_1-2\mu_2,\lambda_2-\lambda_1+\mu_2-\mu_1) \le  &2\nu_1+\nu_2& \scriptstyle \le 2\lambda_1+\lambda_2+2\mu_1+\mu_2 \\ \nonumber
  \scriptstyle\max(\lambda_1+2\lambda_2-2\mu_1-\mu_2,\mu_1+2\mu_2-2\lambda_1-\lambda_2,\lambda_1-\lambda_2+\mu_1-\mu_2) \le & \nu_1+2\nu_2 & \scriptstyle \le  \lambda_1+2\lambda_2+\mu_1+2\mu_2\\ \nonumber
  \scriptstyle \max(\mu_1-\mu_2-2\lambda_1-\lambda_2,\lambda_1-\lambda_2-2\mu_1-\mu_2) \le  &\nu_1-\nu_2&\le \scriptstyle \min(\lambda_1-\lambda_2+\mu_1+2\mu_2,\lambda_1+2\lambda_2+\mu_1-\mu_2) \eea
  or equivalently, making use of the weights $\epsilon_i$ defined in sect \ref{notations}
   \bea  \nonumber \nu_1\ge 0&\qquad & \nu_2\ge 0
  \\ \label{8ineq2}
   \max(\epsilon_3\cdot \lambda +\epsilon_1\cdot\mu , \epsilon_1\cdot\lambda +\epsilon_3\cdot\mu 
  ,\epsilon_2\cdot(\lambda+\mu) ) \le  &\ \ \epsilon_1\cdot\nu &\le \epsilon_1\cdot(\lambda+\mu) 
  \\ \nonumber
  \max(-\epsilon_3\cdot\lambda -\epsilon_1\cdot\mu , -\epsilon_1\cdot\lambda -\epsilon_3\cdot\mu, -\epsilon_2 \cdot(\lambda +\mu)
  \le & -\epsilon_3\cdot \nu   &\le -\epsilon_3\cdot(\lambda +\mu) 
    \\ \nonumber
 \max(-\epsilon_1\cdot\lambda -\epsilon_2\cdot\mu,-\epsilon_2\cdot\lambda -\epsilon_1\cdot\mu)
   \le  &-\epsilon_2\cdot\nu 
  &\le \min( -\epsilon_2\cdot\lambda -\epsilon_3\cdot\mu, -\epsilon_3\cdot\lambda -\epsilon_2\cdot\mu)\,.
  \eea
These  8 inequalities are equivalent to those given by Wesslen \cite{Wesslen}, 
and are written in a way that is explicitly symmetric in the exchange $\lambda\leftrightarrow \mu$
or in the conjugation of representations
($\epsilon_1\leftrightarrow -\epsilon_3,\ \epsilon_2 \leftrightarrow -\epsilon_2$). 
Note  that by adding the second and fourth lines, one gets 
$3\nu_1 \ge (\epsilon_1-\epsilon_2)\cdot \lambda-(\epsilon_1-\epsilon_3)\cdot\mu= \lambda_1-\mu_1-\mu_2
$, etc, 
hence $\nu_1\ge \max( \lambda_1-\mu_1-\mu_2, \mu_1-\lambda_1-\lambda_2,0)$ which was another of Wesslen's inequalities,
clearly not independent of the former inequalities;
likewise $\nu_2\ge  \max(\lambda_2-\mu_1-\mu_2,\mu_2-\lambda_1-\lambda_2,0)$. 
Finally adding the second and third lines yields $\nu_1+\nu_2\le \lambda_1+\lambda_2+\mu_1+\mu_2$, which expresses
that the level 
 of $\nu$ is bounded by   {the level}  of the ``highest highest weight" $\lambda+\mu$ 
 (recall that for SU(3), the level of an irrep is the sum of its Dynkin labels.)
    

\subsection{Multiplicity of $\nu$}
\label{multiplicity}
This subsection is devoted to writing explicit expressions for the multiplicity of $\nu$ in the tensor product $\u\otimes \v$.
We shall need such expressions in our subsequent comparison of multiplicities in 
$\u\otimes \v$ and $\u\otimes \bar\v$.

Given two weights $\lambda=(\lambda_1,\lambda_2)$ and $\mu=(\mu_1,\nu_2)$ of SU(3), 
the above discussion also yields the multiplicity of $(\nu_1,\nu_2)$ 
in $\lambda\otimes \mu$, or equivalently of $1$ in 
$\lambda\otimes \mu\otimes \bar\nu$, with now  $\bar\nu=(\nu_2,\nu_1)$.
It is given by the number of possible values of $\a$, namely the range between the highest 
lower bound and the lowest upper 
bound on $\a$  written in (\ref{ineqa}). 
\bea\label{ineqb} 
{\rm mult}(\lambda,\mu;\nu)= {\rm mult}\!\!\!\!\!&(&\!\!\!\!\!\!\lambda\otimes\mu\otimes \bar\nu\to 1)=  {\rm mult}(\lambda\otimes\mu \to \nu) \\
 \nonumber = 1 + \min\ \!\!\!\!\!\!&(&\!\!\!\!\!\!\lambda_1+2\lambda_2 +\mu_1+\mu_2, \Sigma+\lambda_1+\lambda_2, \lambda_1+2\lambda_2+\mu_1+2\mu_2  -\nu_3) 
 \\ \nonumber
\qquad-\max(\lambda_2+\Sigma, -\mu_2 +\Sigma+ \nu_2+\nu_3\!\!\!\!\!\!&,&\!\!\!\! \Sigma+ \nu_3,\lambda_1+\lambda_2+\mu_1+\mu_2, \lambda_1+2\lambda_2+\mu_2 , 
\lambda_1+\lambda_2 + \nu_2+\nu_3)\,,\eea
with $\Sigma$ and $\nu_3$ given above as functions of $\lambda_1,\lambda_2,\mu_1,\mu_2,\nu_1,\nu_2$. 
Noting that \\ (i) $-\max(a,b,c)=\min(-a,-b,-c)$ and \\ (ii) that $\min(a_i)_{i=1,m}+\min(b_j)_{j=1,n}=
\min(a_i+b_j)_{\genfrac{}{}{0pt}{}{i=1,m} {j=1,n}}$,  and using (\ref{cons-trial}), we find the formula
\bea\label{multnu} \nonumber \mult(\lambda,\mu;\nu) &=& \max (\scriptstyle 0,\  1+\min\{\lambda_1,\  \lambda_2,\  \mu_1,\ \mu_2,\ \nu_1,\ \nu_2,\  \frac{1}{3} ( 2 \lambda_1+\lambda_2+2 \mu_1+\mu_2-2 \nu_1-\nu_2),\ \frac{1}{3} (\lambda_1-\lambda_2+\mu_1+2 \mu_2-\nu_1+\nu_2),\ \\ \nonumber
  & & \scriptstyle \frac{1}{3} (\lambda_1+2 \lambda_2+\mu_1+2 \mu_2-\nu_1-2 \nu_2),\  \frac{1}{3} (2 \lambda_1+\lambda_2-\mu_1+\mu_2+\nu_1-\nu_2),\ \frac{1}{3} (\lambda_1+2  \lambda_2+\mu_1-\mu_2-\nu_1+\nu_2),\ \\ \nonumber
    & & \scriptstyle \frac{1}{3} (2 \lambda_1+\lambda_2-\mu_1-2 \mu_2+\nu_1+2 \nu_2),\ \frac{1}{3} (\lambda_1+2 \lambda_2-2 \mu_1-\mu_2+2\nu_1+\nu_2),\  \frac{1}{3} (-\lambda_1+\lambda_2-\mu_1+\mu_2+\nu_1+2 \nu_2),\ \\ \nonumber
      & & \scriptstyle \frac{1}{3} (\lambda_1-\lambda_2+\mu_1-\mu_2+2 \nu_1+\nu_2),\ \frac{1}{3} (-\lambda_1-2 \lambda_2+2 \mu_1+\mu_2+\nu_1+2 \nu_2),\ \frac{1}{3} (-\lambda_1+\lambda_2+2 \mu_1+\mu_2+\nu_1-\nu_2),\ \\ 
 & & \scriptstyle \scriptstyle  \frac{1}{3} (-2 \lambda_1-\lambda_2+\mu_1+2 \mu_2+2 \nu_1+\nu_2)\})\eea
 which is  made out of the arguments of the inequalities (\ref{systineq}) (up to factors $\frac{1}{3}$) and
which is symmetric under $S_3$ permutations of $(\lambda_1,\lambda_2),(\mu_1,\mu_2), (\nu_2,\nu_1)$ and  complex conjugation 
$(\lambda_1,\mu_1,\nu_2)\leftrightarrow  (\lambda_2,\mu_2,\nu_1)$.
   Alternative expressions making use of the weights $\epsilon_i$ of the fundamental representation read
 \bea\label{multnu2} \nonumber \mult(\lambda,\mu;\nu) &=& \max ( 
 0,\  1+\min\{\lambda_1,\  \lambda_2,\  \mu_1,\ \mu_2,\ \nu_1,\ \nu_2,\  \epsilon_1\cdot (\lambda+\mu-\nu),\ -\epsilon_3\cdot (\lambda+\mu-\nu),
  \\ \nonumber
   \ \epsilon_1\cdot\lambda + \epsilon_2\cdot (\mu-\nu)\!\!\!\!\!
   &, &\!\!\!\!\! -\epsilon_3\cdot\lambda - \epsilon_2\cdot (\mu-\nu),\ 
   \epsilon_1\cdot\lambda + \epsilon_3\cdot (\mu-\nu),\  -\epsilon_3\cdot\lambda - \epsilon_1\cdot (\mu-\nu),
     \\ \nonumber
    \epsilon_3\cdot(\lambda-\nu) +\epsilon_1\cdot\mu \!\!\!\!\!
   &, &\!\!\!\!\!  
            -\epsilon_1\cdot(\lambda-\nu) -\epsilon_3\cdot\mu ,\ \epsilon_2\cdot(\lambda-\nu) +\epsilon_1\cdot\mu,\  -\epsilon_2\cdot(\lambda-\nu) -\epsilon_3\cdot\mu ,  \\ 
       & &
          \epsilon_2\cdot(\lambda+\mu) -\epsilon_3\cdot\nu,\  -\epsilon_2\cdot(\lambda+\mu) +\epsilon_1\cdot\nu
 \})\eea
   or equivalently
   \bea \nonumber 
\mult(\lambda\otimes\mu\otimes \bar\nu\to 1)=\max\Big(0  ,\!\!\!& 1+\min\big(\{\lambda\cdot\alpha_i, \mu\cdot\alpha_i, \bar\nu\cdot\alpha_i\}_{i=1,2,3},\
\epsilon_1\cdot w_I +\epsilon_2\cdot(w_{II}+w_{III}), 
\\ \nonumber   \epsilon_3\cdot w_I +\epsilon_1\cdot(w_{II}+w_{III}),\!\!\!\!\!\!\!\!& 
-\epsilon_1\cdot w_I -\epsilon_3\cdot(w_{II}+w_{III}), -\epsilon_3\cdot w_I -\epsilon_2\cdot(w_{II}+w_{III})\big)\Big)
\eea
where $w_I$ denotes one of the three weights $\lambda, \mu, \bar\nu$ and $w_{II},w_{III}$ the two others.
This expression is explicitly symmetric under $S_3$ permutations of $\lambda, \mu, \bar\nu$, and under complex
conjugation where $\epsilon_1\leftrightarrow -\epsilon_3$ and 
$\epsilon_2\leftrightarrow -\epsilon_2$. 

\begin{figure}\begin{center}
\includegraphics[width=.45\textwidth]{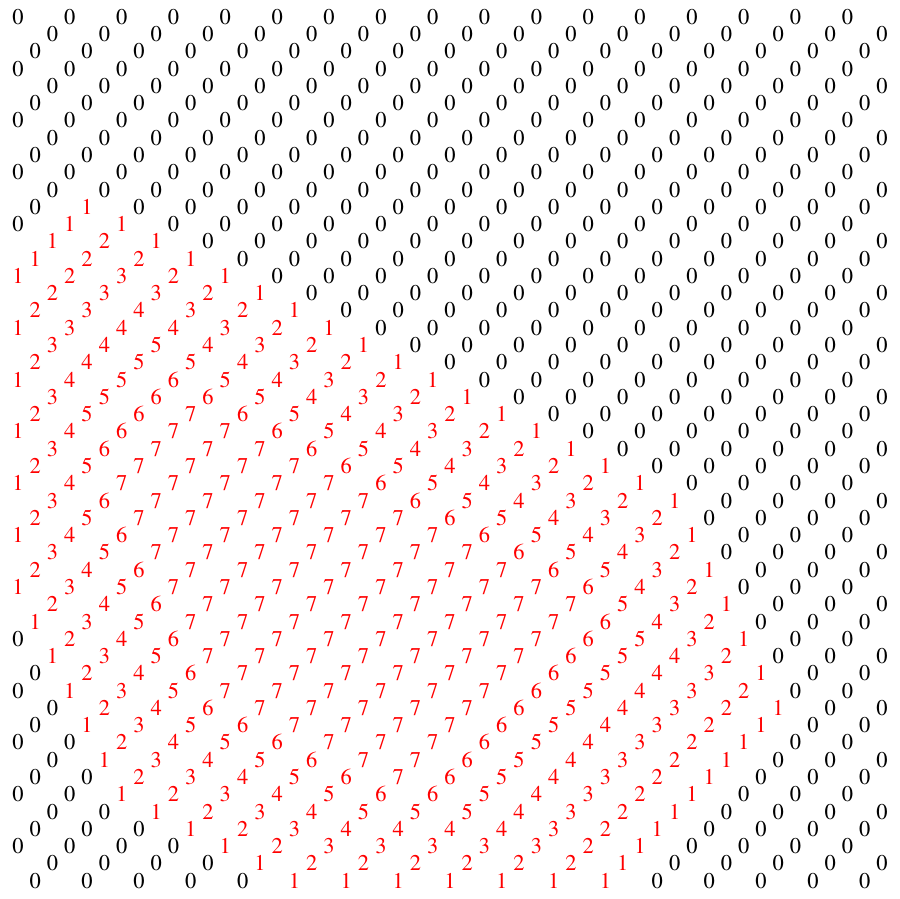}
\end{center}
\caption{\footnotesize Multiplicities for $\lambda_1=21,\lambda_2=6,\mu_1=17,\mu_2=16$, in a $\nu_1,\nu_2$ 
rectangular plot}
 \label{multip}
\end{figure}

Note that one recovers the ``well known" fact that if any of the Dynkin labels $\lambda_1,\  \lambda_2,\  \mu_1,\ \mu_2,\ \nu_1,\ \nu_2$ vanishes, 
 the multiplicity is 0 or 1: this is Pieri's rule, see also (\ref{maximalmultiplicity}).
Either of these formulae enables one to set up the plot of multiplicities, see Fig. \ref{multip} for an example.

 Our last purpose, in this section, is to express in a simpler way the formula (\ref{multnu}) giving the multiplicity. 
\\
Call $S_1 = \lambda_1+\mu_1+\nu_2$, $S_2 = \lambda_2+ \mu_2+ \nu_1$.
There are three cases :\\
Case 1.   $S_1 = S_2$ (call it $S$).  The previous expression (\ref{multnu}) simplifies immediately and reads 
\be\label{If1}\text{If} \quad S_1 = S_2,  \quad \mult(\lambda, \mu; \nu) = \max(0,1+\min(\lambda_1,\lambda_2,\mu_1,\mu_2,\nu_1,\nu_2,L,M,N))\ee
 where $L = S- (\lambda_1+\lambda_2)$, $M = S-(\mu_1+\mu_2)$ and $N = S - (\nu_1 + \nu_2)$.
Notice that the multiplicity does not vanish if and only if  $L$, $M$ and $N$ are non-negative.\\
Case 2.  $S_1 > S_2$.   Call $s = (S_1-S_2)/3$. Notice that $s$ should be an integer 
since $S_1-S_2= \tau(\lambda)+\tau(\mu) -\tau(\nu)\, \mod 3$. 
Equation (\ref{multnu}), together with $S_1 > S_2$,  implies that\footnote{notice that the three integers  $\lambda_1 - s,  \mu_1 - s, \nu_2-s$ are non-negative. The fact that $\lambda_1 - s > 0$ is equivalent to the third Wesslen inequality (lhs of eq 14);  the two other inequalities are obtained from the first by a symmetry argument.} 
\be\label{If2}\mult((\lambda_1 - s, \lambda_2), (\mu_1 - s, \mu_2) ; (\nu_1, \nu_2-s)) = \mult((\lambda_1, \lambda_2), (\mu_1, \mu_2) ; (\nu_1, \nu_2))\,. \ee
Proof:  Using eq. (\ref{multnu}), one sees that both multiplicities simplify to 
$\max(0, 1 +  \min(\lambda_2,\mu_2,\nu_1,\frac{1}{3}(\lambda_1+2 \lambda_2+\mu_1+2 \mu_2-\nu_1-2 \nu_2),\frac{1}{3} (2 \lambda_1+\lambda_2-\mu_1+\mu_2+\nu_1-\nu_2),\frac{1}{3} (-\lambda_1+\lambda_2+2 \mu_1+\mu_2+\nu_1-\nu_2),\frac{1}{3} (\lambda_1+2 \lambda_2-2 \mu_1-\mu_2+2 \nu_1+\nu_2),\frac{1}{3} (-2 \lambda_1-\lambda_2+\mu_1+2 (\mu_2+\nu_1)+\nu_2),\frac{1}{3} (-\lambda_1+\lambda_2-\mu_1+\mu_2+\nu_1+2 \nu_2)))$, hence the result.
By construction, the sums  $(\lambda_1 - s) + (\mu_1 - s) + (\nu_2-s)$ and  $\lambda_2+ \mu_2+ \nu_1$ are equal, so that one can now use the previous simplified result (Case 1) to calculate the multiplicity.\\
Case 3. $S_1 < S_2$. Call  $s = (S_2-S_1)/3$ and proceed like in case 2 but shift by $s$ the components $\lambda_2, \mu_2, \nu_1$ instead.\\[5pt]
In a nutshell: To every triple of  weights labelling a space of intertwiners, one can associate an integer $s$, as defined above.
By an appropriate shift of three of the components of those three weights (see above) one can always build another triple of weights for which $s$ vanishes.  The dimension of the space of intertwiners is the same for both triples.  In order to determine this multiplicity, one therefore uses the triple for which $s=0$. The multiplicity is obtained from the components of its three weights by a very simple formula (cf. case 1 above).\\
Example.
Let us calculate $\mult((21, 6),(17, 16); (12, 8)) = \dim ((21, 6) \otimes (17, 16) \otimes (8, 12)  \rightarrow 1)$. We have $(S_1,S_2) = (46,34)$, and $s = (S_1-S_2)/3 = 12/3 = 4$.
So, we consider instead $(17,6)\otimes (13,16)\otimes (4,12)   \rightarrow 1$  since $\mult((21, 6),(17, 16); (12, 8)) = \mult(17,6), (13,16); (12,4))$.
The two sums are now equal to $S = 17+13+4=6+16+12 = 34$. One then gets $L = 34-(17+6) = 11$, $M = 34 - (13+16)=5$, $N = 34 - (12+4)=18$ and one obtains immediately the multiplicity $1 + \min(17,6,13,16,12,4,11,5,18)=5$.\\
Our expression (\ref{multnu}) for the multiplicity is fully explicit, but it appears that, in the case of SU(3),  equivalent formulae have been discovered and rediscovered several times in the literature, and even generalized to the case of fusion multiplicities, \ie for the affine algebra su(3) at some finite level.  The results of the present section are therefore hardly new. For other presentations, the reader may refer for instance to \cite{BMW}, or to the thesis  \cite{Suciu}. In the latter case, the results are proved in the framework of TQFT for SU(3), using a graphical calculus that we don't use in the present article.\\

\subsection{Convexity and tensor polygons} 
\label{convexity}
\def\R{\mathbb{R}}

Recall that in $\R^d$, the points $x_i$ satisfying a set of inequalities $H_j(x_i)\ge 0$, where $H_j=0$,  
$j=1,2,\cdots p$,
 are equations of $p$ hyperplanes, form a convex polytope, bounded or not bounded. 
 
 Here, in SU(3), we may apply this argument either to the set of points $(\nu,{\a})$ which live in $\R^{2+1}$ and are subject to
 inequalities (\ref{ineqa}), or to 
 points ${\a}$ (in the finite segment (\ref{ineqb})), or to the resulting $\nu\in \R^2$, cf (\ref{systineq}) or (\ref{8ineq}). 
  Notice that the fact that $\nu$ should obey the triality constraint  (\ref{cons-trial})  
 means that $\nu$ belongs to 
 the root lattice $Q$ shifted by some weight of $\lambda\otimes \mu$, 
  for example the ``highest highest weight" $\lambda+\mu$
 $$ \nu \in Q + \lambda+\mu\,.$$ 
 Moreover it is clear that the number of points, in either of the previous contexts, either $(\nu,{\a})$, or ${\a}$, or $\nu$, is finite, 
  for given $\lambda$ and $\mu$.
 Thus the previous statement should be rephrased as\\
{ \sl $\nu$ belongs to a convex bounded polytope drawn on the lattice $  Q + \lambda+\mu$, thus 
reaches all integer points interior to or on the
 boundary of that polytope.}\\
 This is a particular case of a convexity property proved by Knutson and Tao for general GL($N$).

\begin{figure}
   \mbox{ 
  \setlength{\unitlength}{1.3pt}
\begin{picture}(100,60)
\put (180,0){}
\put(205,18){$H=V_1$}\put(200,18){\vector(1,-2){25}}\put(210,0){$\scriptstyle \min(\q,\s)\, (1,-2)$} 
\put(225,-32){\vector(-1,-1){25}}\put(220,-45){$\scriptstyle |\min(\q-\s,\r)|\, (-1,-1)$}\put(225,-35){$V_2$}
\put(210,-57){$V_3$}
\put(200,-57){\vector(-3,0){20}}
\put(210,-70){\vector(-2,1){20}}
\put(210,-70){$\scriptstyle \max[\min(\p,\q,\r,\s,\r+\s-\q),0] \, (-3,0)$}
\put(180,-57){\vector(-2,1){20}}
\put(145,-63){\vector(2,1){20}}
\put(170,-66){$V_4$}
\put(15,-73){$\scriptstyle\begin{cases}\scriptstyle \min(\p-\s,\r-\q)&\scriptstyle {\rm if\ } \r> \q\\
\scriptstyle |\min(\q-\r,\s)|&\scriptstyle{\rm if\ }\q\ge r \end{cases} (-2,1)$}
\put(125,-50){$h=V_5$}
\put(160,-47){\vector(-1,2){20}}\put(5,-35)
{$\scriptstyle |\min(\p-\r-\s,0)+\min(\q,\r)| 
 (-1,2)$}
\put(128,-13){$V_6$}
\put(140,-7){\vector(0,3){25}}\put(20,2){$\scriptstyle \max[\min(\q,\r+\s-\p),0] \, (0,3)$}
\put(128,18){$V_7$}
\put(140,18){\vector(1,1){20}}\put(65,30){$\scriptstyle |\min(\p-\r,\s)|\, (1,1)$}
\put(160,38){\vector(2,-1){40}}\put(180,30){$\scriptstyle \r 
\, (2,-1)$}
\put(155,40){$V_8$}
\end{picture}}
\vskip45mm
\caption{The oriented sides of {a} tensor polygon (with the assumption that  $\p\ge \max(\q,\r,\s)$). 
\footnotesize(We have used orthogonal coordinate axes,  although it would be more correct to have them at a $\pi/3$ angle.)
}\label{wessl-octo}
\end{figure}
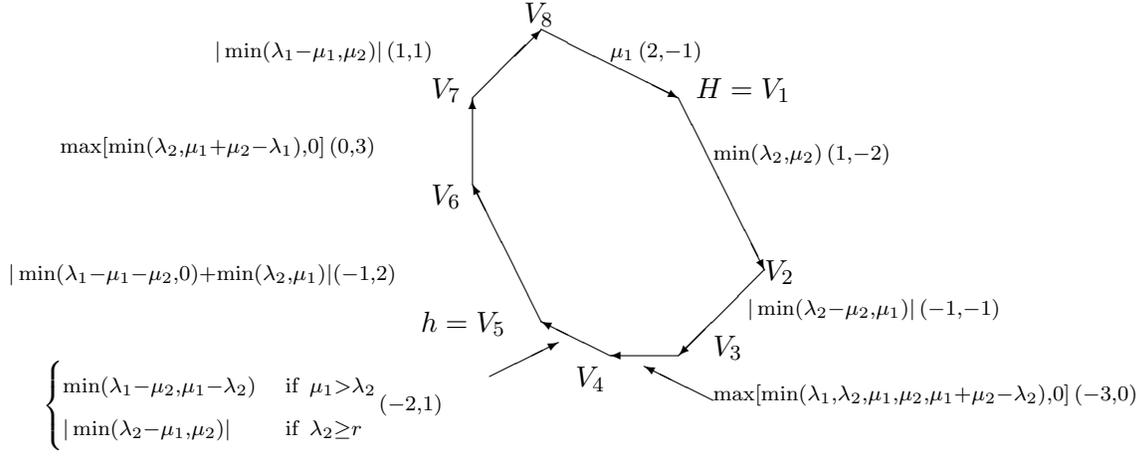

This convex polygon has been charted by Wesslen \cite{Wesslen}: it has
 $v\le 8$  vertices   as depicted in Fig. \ref{wessl-octo}, where we have assumed that $\p= \max(\p,\q,\r,\s)$. 
 Two vertices play a special r\^ole: $H=(\p+\r,\q+\s)$ is the ``highest highest weight" (h.h.w.), 
 while  the ``lowest highest weight" $h$ is defined by
\be h=\begin{cases}  (\r+\s-\p-\q,\p-\s) &{\rm if\ } 0\le \p-\s\le \r-\q \\ 
                                                           (\p+\q-\r-\s, \r-\q) & {\rm if \ } 0\le \r-\q\le \p-\s \\ 
                                                            (\p-\s,\q-\r) &{\rm if\ } \r-\q<0\,.
   \end{cases}\ee

  The coordinates of the other vertices may be readily computed from the data of Fig. \ref{wessl-octo}, which gives 
  the components of the edge vectors in the $(\omega_1,\omega_2)$ basis. Note that some edges may 
  ``degenerate" (\ie be of length 0) and the octogon reduce to an heptagon, an hexagon, \dots , or a single point.
  
 \def\CC{{\cal C}}
 From the formula (\ref{multnu}) for the multiplicity of a weight $\nu$, it follows that the polygons of increasing multiplicity
 have a ``matriochka" pattern. Let $\CC_m$ the set of $\nu$'s of multiplicity $\ge m$. If $H_j$, $j=1,\cdots 18$ denote the
 18 arguments of $\min$ in  (\ref{multnu}), the points in $\CC_{m+1}$ satisfy $H_j \ge m$, hence {\it a fortiori} 
 $H_j\ge m-1$, thus  $\CC_{m+1} \subset \CC_m$.

 Call $P^{(m)}$ the set of all $\nu$'s belonging to the  boundary of the polygon determined by $\CC_{m}$.
 More precisely, if we call $P^{(1)}(\lambda,\mu)$ the outermost tensor polygon, $P^{(2)}(\lambda,\mu), \cdots,
 P^{(\rm mult_{\rm max})}(\lambda,\mu)$ the successive ones, one easily sees that they obey
 \be\label{matriochK} P^{(m+1)}(\lambda,\mu)=P^{(1)}(\lambda-m\rho,\mu-m\rho)+m\rho\ee
 where $\rho=\omega_1+\omega_2 =(1,1)$ is the Weyl vector.
 This implies, if $m < \mult_{max}$, that $P^{(m)} = \CC_{m+1} \setminus \CC_{m} $ (set difference).  
 In other words,  if $m < \mult_{max}$, all weights of multiplicity $m$ lie on the sides of the polygon $P^{(m)}$.
 See for example Fig. \ref{protrudedpolyhedron} or \ref{multip}. 
 
 %

\subsection{From honeycombs to hives} 
\label{honeycomb-hive}
\begin{figure}[htbp]
\setlength{\unitlength}{1.7pt}
{\begin{picture}(100,50)(-30,0)
                           \put (70,45){$\scriptstyle 0$}           
                        			   \put(99,50){\circle*{2}} 
               		   \put(72,7){\circle*{2}}  \put(125,7){\circle*{2}}    
		             \multiput(72,7)(6,0){10}{\line(-1,0){1}}    
          \multiput(100,50)(-3,-5){29}{\line(-1,0){1}} \multiput(100,50)(3,-5){29}{\line(-1,0){1}} 
          \multiput(72,7)(2.8,-5){20}{\line(-1,0){1}}      \multiput(125,7)(-2.6,-5){20}{\line(-1,0){1}}  
   \put(44,-42){\circle*{2}}               \put(99,-42){\circle*{2}}                         \put(154,-42){\circle*{2}}  
                        \multiput(44,-42)(6,0){20}{\line(-1,0){1}}    
     \multiput(44,-42)(3.2,-5){10}{\line(-1,0){1}}                                \multiput(154,-42)(-3.2,-5){10}{\line(-1,0){1}}  
                   \multiput(16,-90)(6,0){29}{\line(-1,0){1}}    
                          \put(14,-90){\circle*{2}} 
                           \put(75,-90){\circle*{2}}   \put(125,-90){\circle*{2}}  \put(186,-90){\circle*{2}}  
                             \put(100,20){\line(1,1){25}}\put(100,20){\line(-1,1){25}} \put (127,45){$\scriptstyle \mu_1+\mu_2$}
                                                                     \put(100,20){\line(0,-1){25}}    \put(90,5){${\scriptstyle -\mu_1-\mu_2}$}
                                        \put(68,-20){${\scriptstyle  -\lambda_1-2\lambda_2 + \a} $}    
                                          \put(105,-18){${\scriptstyle \lambda_1+2\lambda_2+\mu_1}$}    \put(108,-25){${\scriptstyle   {+\mu_2- \a}} $}                           
 \put (43,-5){$\scriptstyle \lambda_2$}  \put(75,-30){\line(-1,1){25}}\put(100,-5){\line(-1,-1){25}}\put(100,-5){\line(1,-1){25}}\put(125,-30){\line(1,1){25}}
 			\put(60,-45){${\scriptstyle  \lambda_1+\lambda_2  {- \a}}$}\put(110,-41){${\scriptstyle   -\lambda_1-2\lambda_2}$}\put(110,-48){${\scriptstyle  -\mu_1-2\mu_2} {\scriptstyle{+ \a}}$}
                                                       \put(75,-30){\line(0,-1){25}} \put(125,-30){\line(0,-1){25}} \put (153,-5){$ \scriptstyle \mu_2$}
 \put (5,-55){$\scriptstyle\lambda_1+\lambda_2$}   \put(25,-55){\line(1,-1){25}}    \put(50,-70){${\scriptstyle \Sigma -\lambda_1-\lambda_2}$}
  \put(83,-70){${\scriptstyle \a-\Sigma}$}\put(100,-70){${\scriptstyle \nu_2+\nu_3+\Sigma}{\scriptstyle - \a}$} \put(135,-70){${ \scriptstyle \nu_3} $} 
  \put(75,-55){\line(-1,-1){25}}\put(75,-55){\line(1,-1){25}}\put(125,-55){\line(-1,-1){25}}\put(125,-55){\line(1,-1){25}}  \put(175,-55){\line(-1,-1){25}} \put (178,-55){$\scriptstyle 0$}
                 \put(50,-80){\line(0,-1){25}} \put(100,-80){\line(0,-1){25}}\put(150,-80){\line(0,-1){25}} 
                  \put(30,-115){$\scriptstyle -\Sigma\equiv -\nu_1-\nu_2-\nu_3$}         \put(90,-115){$\scriptstyle -\nu_2-\nu_3$}  \put(143,-115){$\scriptstyle
                  -\nu_3$}       
                   \put(84,55){$\blue{\lambda_1+2\lambda_2}$}
                 \put(42,7){$\blue{\lambda_1+2\lambda_2}$}  \put(130,7){$\blue{\lambda_1+2\lambda_2+\mu_1+\mu_2}$}
                   \put(17,-42){$\blue{\lambda_1+\lambda_2}$}\put(93,-44){$\red{\a}$}\put(156,-42){$\blue{\lambda_1+2\lambda_2+\mu_1+2\mu_2}$}
                    \put (6,-99){$\blue{0}$}    
                     \put(52,-99){$\blue{\lambda_1+2\lambda_2+\mu_1}$}  \put(48,-109){$\blue{+2\mu_2-\nu_2-2\nu_3}$}   \put(102,-99){$\blue{\lambda_1+2\lambda_2+\mu_1}$}  \put(105,-109){$\blue{+2\mu_2-\nu_3}$} \put(186,-99){$\blue{\lambda_1+2\lambda_2+\mu_1+2\mu_2}$} 
\end{picture}}
\vskip7cm
\caption{ Constructing the KT hive from the KT honeycomb}
\label{honey2hive1}
\end{figure}
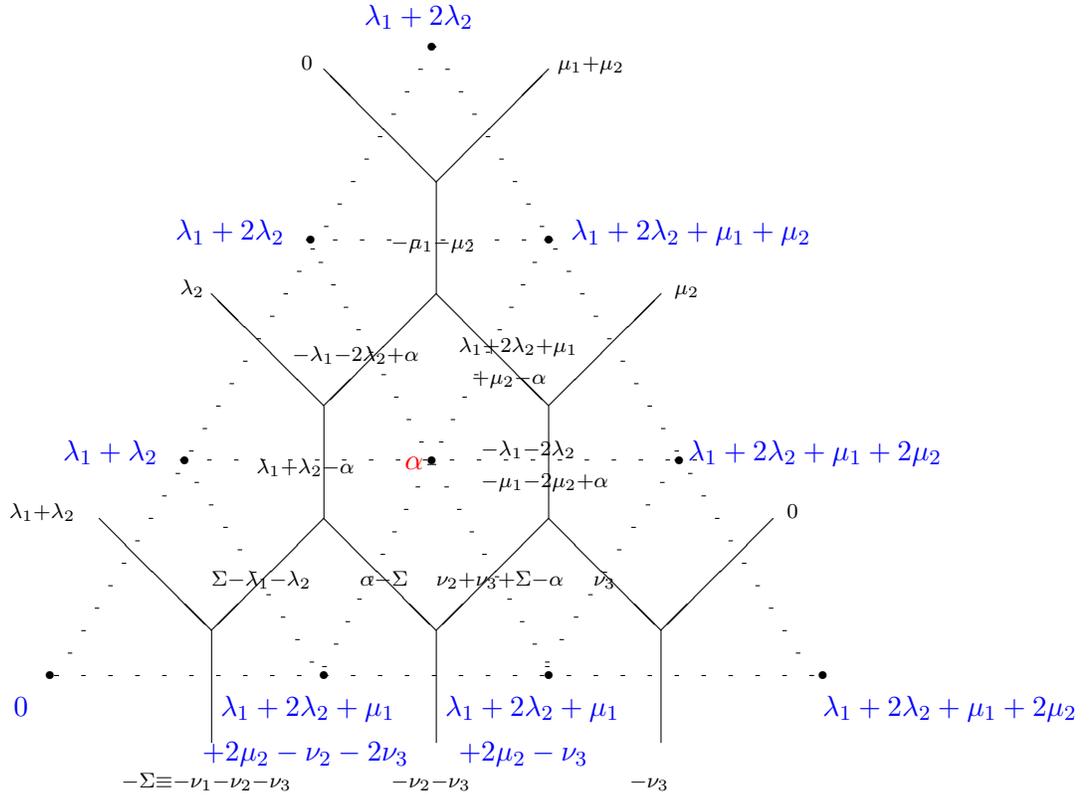

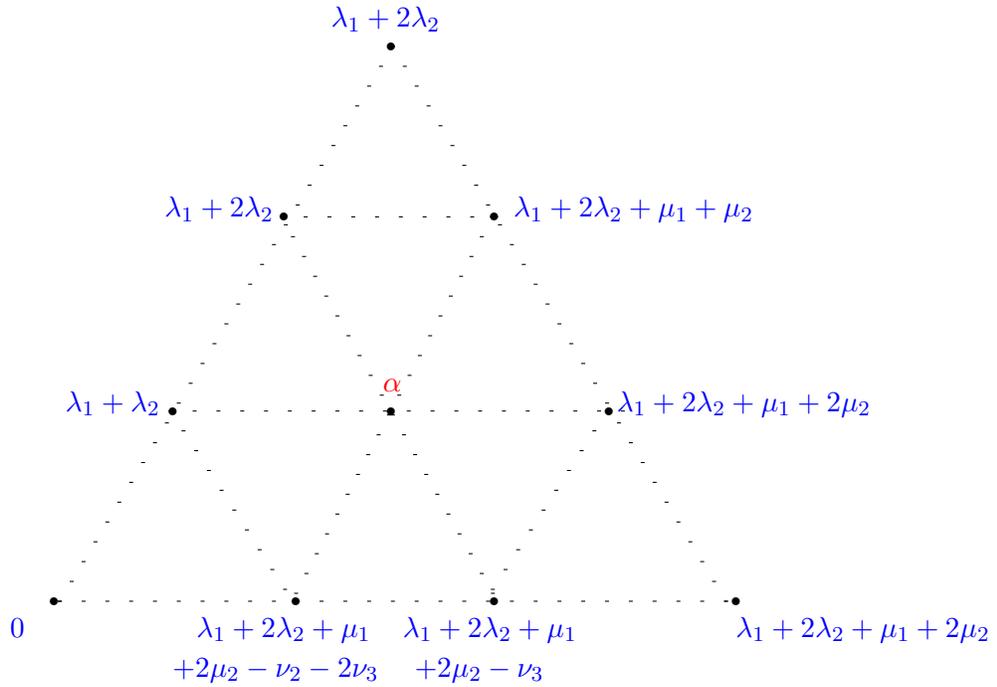
\begin{figure}[htbp]
\setlength{\unitlength}{1.5pt}
{\begin{picture}(100,50)(-50,0)
                        			   \put(99,50){\circle*{2}} 
               		   \put(72,7){\circle*{2}}  \put(125,7){\circle*{2}}    
		             \multiput(72,7)(6,0){10}{\line(-1,0){1}}    
          \multiput(100,50)(-3,-5){29}{\line(-1,0){1}} \multiput(100,50)(3,-5){29}{\line(-1,0){1}} 
          \multiput(72,7)(2.8,-5){20}{\line(-1,0){1}}      \multiput(125,7)(-2.6,-5){20}{\line(-1,0){1}}  
   \put(44,-42){\circle*{2}}               \put(99,-42){\circle*{2}}                         \put(154,-42){\circle*{2}}  
                        \multiput(44,-42)(6,0){20}{\line(-1,0){1}}    
     \multiput(44,-42)(3.2,-5){10}{\line(-1,0){1}}                                \multiput(154,-42)(-3.2,-5){10}{\line(-1,0){1}}  
                   \multiput(16,-90)(6,0){29}{\line(-1,0){1}}    
                          \put(14,-90){\circle*{2}}
                           \put(75,-90){\circle*{2}}   \put(125,-90){\circle*{2}}  \put(186,-90){\circle*{2}}  
                   \put(84,55){$\blue{\lambda_1+2\lambda_2}$}
                 \put(42,7){$\blue{\lambda_1+2\lambda_2}$}  \put(130,7){$\blue{\lambda_1+2\lambda_2+\mu_1+\mu_2}$}
                   \put(17,-42){$\blue{\lambda_1+\lambda_2}$}\put(97,-37){$\red{\a}$}\put(156,-42){$\blue{\lambda_1+2\lambda_2+\mu_1+2\mu_2}$}
                    \put (3,-99){$\blue{0}$}    
                     \put(50,-99){$\blue{\lambda_1+2\lambda_2+\mu_1}$}  \put(44,-109){$\blue{+2\mu_2-\nu_2-2\nu_3}$}   \put(102,-99){$\blue{\lambda_1+2\lambda_2+\mu_1}$}  \put(105,-109){$\blue{+2\mu_2-\nu_3}$} \put(186,-99){$\blue{\lambda_1+2\lambda_2+\mu_1+2\mu_2}$} 
\end{picture}}
\vskip6cm
\caption{The KT hive  in SU(3). }
\label{honey2hive2}
\end{figure}

From the honeycomb picture, one may go the dual ``hive" picture, a triangular pattern of size 
$3\times 3 \times 3$ in the present case of SU(3) (or GL(3)). Each vertex of the hive carries an integer obtained 
from the honeycomb  by the following rule: by convention\footnote{and there are clearly two other possible conventions}  
the lower left corner is assigned 0, and along each upward triangle travelled  clockwise, 
each vertex is obtained from the value of the previous one by adding the honeycomb line which separates them.
In that way, one finds that external vertices carry $(0, \lambda_1+\lambda_2, \lambda_1+2\lambda_2, \lambda_1+2\lambda_2,
 \lambda_1+2\lambda_2+\mu_1+\mu_2,  \lambda_1+2\lambda_2+\mu_1+2\mu_2, \lambda_1+2\lambda_2+\mu_1+2\mu_2,
 \lambda_1+2\lambda_2+\mu_1+2\mu_2-\nu_3,  \lambda_1+2\lambda_2+\mu_1+2\mu_2-\nu_2-2\nu_3, \lambda_1+2\lambda_2+\mu_1+2\mu_2-\nu_1-2\nu_2-3\nu_3
 \equiv 0)$, thanks to (\ref{zz}). See Fig. \ref{honey2hive1} and \ref{honey2hive2}.

  \begin{minipage}[t]{0.50\textwidth}
In the hive picture, the inequalities (\ref{ineqa}) express that for each rhombus made of two adjacent equilateral triangles, 
the sum of weights on the short diagonal is {\bf larger} than or equal to that along the long one.
(In (\ref{ineqa}), the three successive lines correspond to rhombi with a short diagonal respectively in $-$, $\backslash$ 
and $\slash$ directions.)\end{minipage} \hfill
\qquad\begin{minipage}[t]{0.5\textwidth}\vspace{-10pt}
\centering\includegraphics[width=0.5\textwidth]{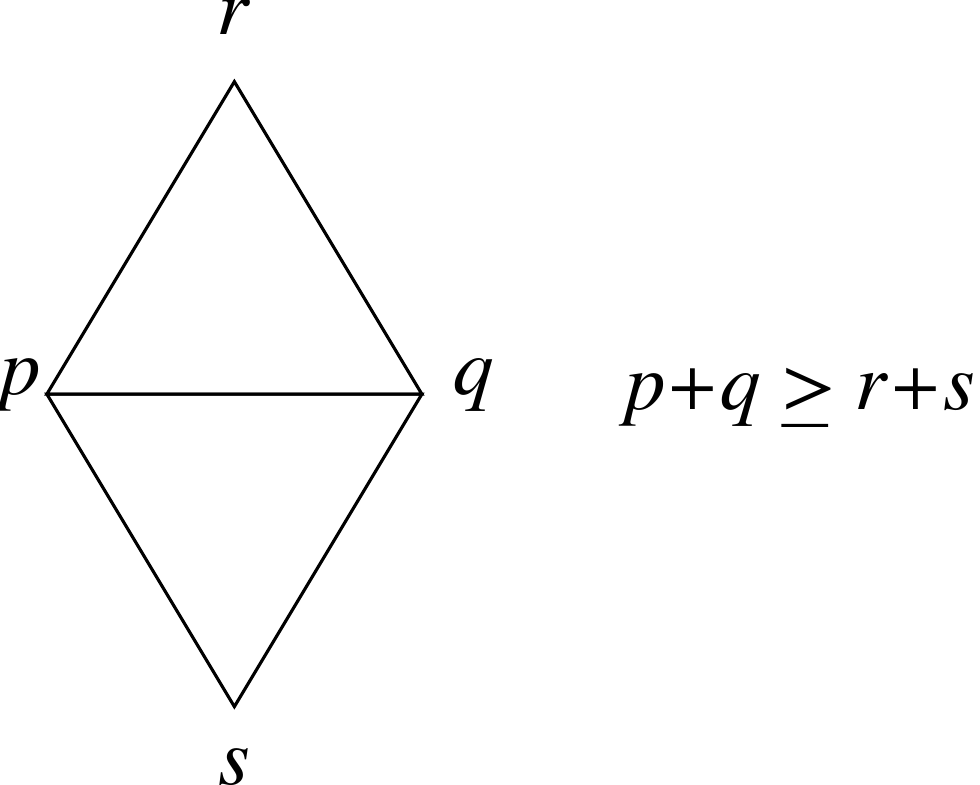}
 \end{minipage}

\def\ot{\inv{3}}
\subsection{Bounds on $\a$} 
 {}From  the convexity argument of sect. \ref{convexity},
 we know that all values of $\a$ in an  interval 
  \be \a_{\rm min}(\lambda,\mu;\nu)
  \le \a \le \a_{\rm max}(\lambda,\mu;\nu)  \,. \ee
satisfy all inequalities (\ref{ineqa}).  Here, we 
make more explicit the expressions of the bounds $\a_{\rm max}(\lambda,\mu;\nu)$ and $ \a_{\rm min}(\lambda,\mu;\nu)$
 of that interval.\\
 For given $\lambda,\mu,\nu$,  the upper bound on $\a$, coming from the inequalities (\ref{ineqa}),  reads
   \bea \label{supa}
   \a_{\rm max}(\lambda,\mu;\nu)  &\!\!=\!\!& 
      \min \Big( \lambda_1+2 \lambda_2+\mu_1+\mu_2, \Sigma+\lambda_1+\lambda_2, \tau(\lambda)+\tau(\mu) -\nu_3    \Big)
    \\
&\!\!=\!\!& \nonumber
   \lambda_1+2 \lambda_2+\mu_1+\mu_2+ \frac{1}{3} {\min}\Big(0,3\lambda_1-\tau(\bar\lambda)-\tau(\bar\mu)+\tau(\bar\nu),
3\mu_2-\tau(\lambda) -\tau(\mu)+\tau(\nu)\Big)\,.
  \eea
 On the other hand $\a_{\rm min}(\lambda,\mu;\nu)$ is determined by 
$$\a_{\rm max}(\lambda,\mu;\nu) -\a_{\rm min}(\lambda,\mu;\nu)=\mult(\lambda,\mu;\nu) -1\,,$$ 
 with the  multiplicity given by one of the  formulae of sect. \ref{multiplicity}.

Given a pair $(\nu,\a)$ satisfying the inequalities (\ref{ineqa}), we may define its multiplicity index by 
\be\label{mult-ind}m(\nu,\a)= \a-\a_{\rm min}(\lambda,\mu;\nu) +1\,.\ee
Clearly $m$ takes values in $\{1,\cdots, \mult(\lambda,\mu;\nu) \}$. We shall make use of that notion 
in the following section.


\subsection{A piecewise linear map $(\nu,\a)\mapsto(\nu',\a')$}
\label{piecewiselinearmap}
We now turn to the issue of writing a mapping between h.w. of irreps appearing in $\lambda\otimes \mu$ and in 
$\lambda\otimes\bar\mu$. In this section, for the sake of simplicity of notations (and to avoid an overflow of min and max),
we assume that $\lambda_1$ is the supremum of all Dynkin labels of $\lambda$ and $\mu$\footnote{Otherwise, we assume 
that we first swap $\lambda$ and $\mu$ and/or conjugate both of them, which does not affect the set of multiplicities.}
$$ \lambda_1 =\max(\lambda_1,\lambda_2,\mu_1,\mu_2)\,.$$
We shall show below that there exists a piecewise linear mapping acting  on pairs $(\nu,a)$
\be (\nu,\a) \mapsto (\nu',\a')\ee
with 
\bea \nu' &=&\begin{cases} 
(\nu_1,\nu_2)+(\mu_2-\mu_1)(1,-1) & {\rm if\ }  2\nu_1+\nu_2> 5\lambda_1+4\lambda_2-\mu_1-2\mu_2{-3(m(\nu,\a)-1)}\cr
  & {\rm or\ } \nu_1+\nu_2> 2\lambda_1+\lambda_2 {-(m(\nu,\a)-1)} \\
(2\lambda_1+\lambda_2-\nu_1-\nu_2,\nu_2) & {\rm otherwise }\end{cases} \cr
\label{transfo}
\a'&=&   
\a -\a_{\rm min}(\lambda,\mu;\nu) + \a_{\rm min}(\lambda,\bar\mu;\nu')  \eea
where the latter equation expresses that the multiplicity indices of $(\nu,\a)$ and of $(\nu',\a')$ 
as defined in (\ref{mult-ind}) are the same.

We stress that a given h.w. $\nu$ coming in $\lambda\otimes\mu$ with some multiplicity may be  sent 
onto {\it several} distinct 
h.w. $\nu'$, depending on its multiplicity index (or equivalently on the value of ${\a}$). The map (\ref{transfo}) is, however, one-to-one 
between the sets of allowed $(\nu,{\a})$ and $(\nu',{\a}')$. 
Therefore it maps all allowed weights $\nu$ appearing in $\lambda\otimes\mu$, with their multiplicity,
onto all allowed weights $\nu'$ appearing in $\lambda\otimes\bar\mu$, with their multiplicity. This is thus a new proof of
Theorem 2, namely
that the lists of multiplicities in $\lambda\otimes\mu$ and $\lambda\otimes\bar\mu$ are the same
up to permutations. 

\def\u{\lambda} \def\v{\mu}\def\w{\nu}

\begin{figure}\begin{center}
\includegraphics[width=.45\textwidth]{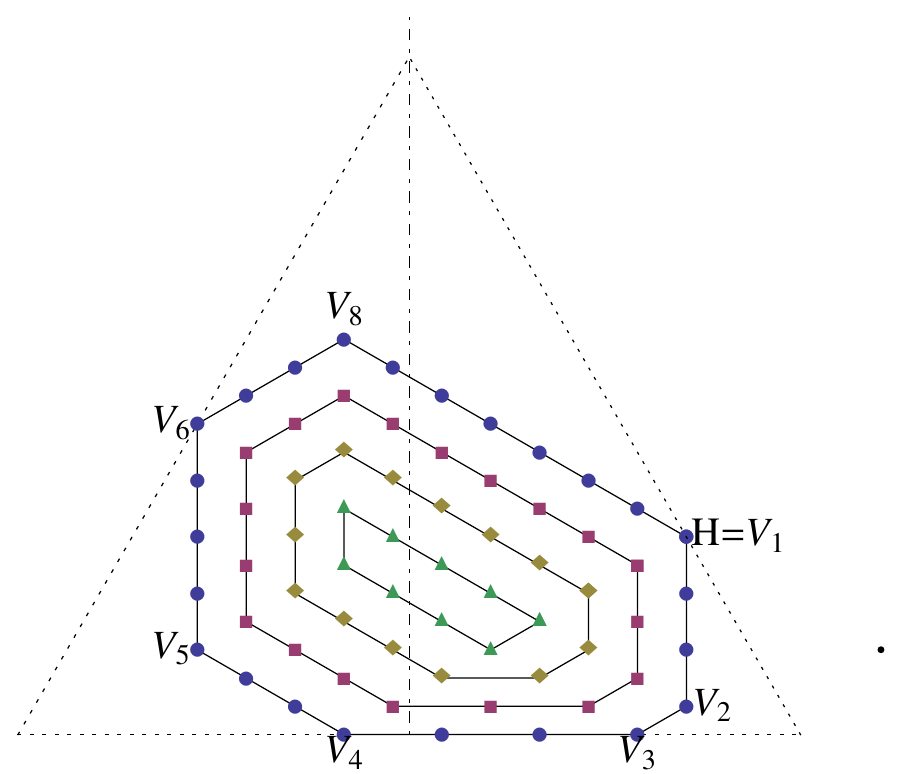}\qquad\includegraphics[width=.45\textwidth]{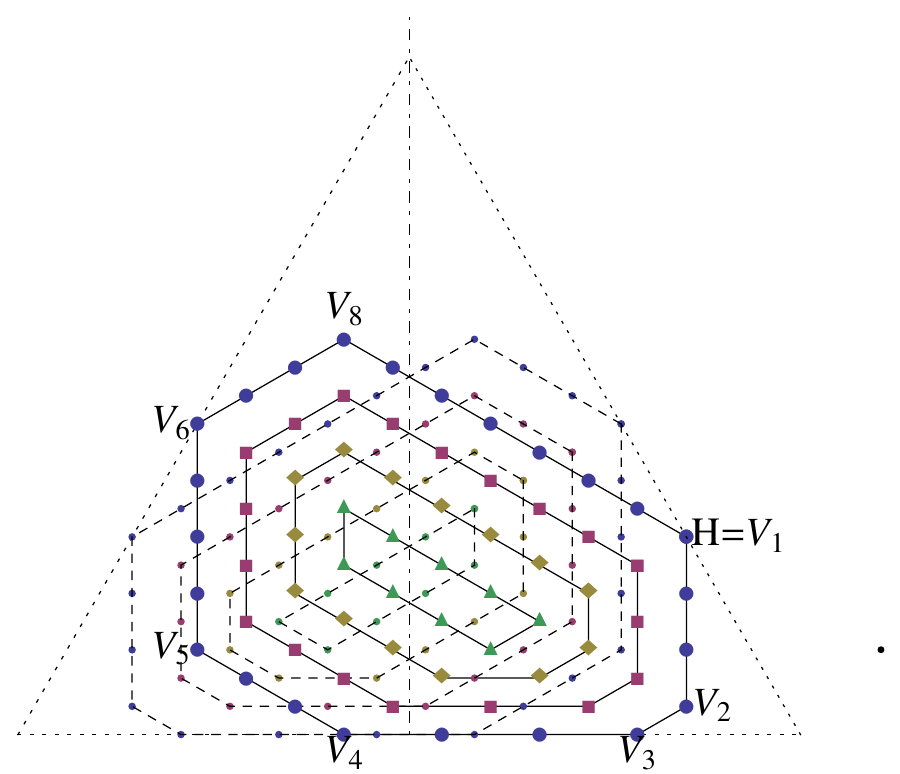}
\end{center}
\caption{\footnotesize The two polygons for $(p,q,r,s)=(10,4,7,3)$ and $(10,4,3,7)$, with the vertical reflection axis. }
 \label{tens-poly-a}
\end{figure}

\be\label{t2}(\x,\y)\mapsto (\x',\y')=\tr_1(\x,\y)=(2\p+\q-\x-\y,\y)\,.\ee
This may be recast into $2\x'+\y'=2(2\p+\q)-(2\x+\y)$ and $\y'=\y$, and as $2\x+\y=\langle 2\alpha_1+\alpha_2,\nu\rangle$
while $\y=\langle \alpha_2,\nu\rangle$, and $\langle 2\alpha_1+\alpha_2, \alpha_2\rangle =0$, 
the  transformation $\w'=\tr_1(\w)$ is 
geometrically  an orthogonal  reflection in the line of equation 
$\langle 2\alpha_1+\alpha_2,\nu\rangle =2\p+\q$, \ie  in the line orthogonal to $\epsilon_1=\omega_1$ passing 
through $\lambda$, 
see Fig.\,\ref{tens-poly-a}. This explains why $\tr_1$ is involutive (for a given $\u$): $\tr_1\circ \tr_1={id}$. 

One readily checks that all $(\x',\y')$ have the right triality to appear in the rhs of $\u\otimes \bar \v$. 
If $\w=(\x,\y)$ has the triality of $\u\otimes \v$
\be\nonumber \tau(\w)\equiv \x-\y \equiv \p+\r-\q-\s\ \mod 3\ee
then
\bea\nonumber \tau(\w')&=& \x'+2\y'\equiv 2\p+\q-\x+\y  \equiv 2\p+\q-(\p-\q+\r-\s) \\ 
&\equiv& \p+2\q+\s-\r \equiv \p+\s-\q-\r\equiv \tau(\u\otimes \bar \v)\ \mod 3\eea
as it should. 

We show below that for $\u$ and $\v$ ``deep enough" in the fundamental Weyl chamber, in a sense to be made precise, 
the transformation $\tr_1$ maps all $\w$ appearing in $\u\otimes \v$ onto all $\w'$ appearing in $\u\otimes\bar\v$
while preserving their multiplicity. For generic $\u$ and $\v$,
however, the transformation $\tr_1$ may map some of the points 
inside (or on the boundary of) the polygon $P^{(1)}(\lambda,\mu)$ {\it outside} the polygon $P^{(1)}(\lambda,\bar\mu)$, 
in other words, the transformed weight $\nu'$ violates some of the inequalities (\ref{8ineq2}), in which 
case we have to modify the transformation. This is apparent in Fig. \ref{tenspolyu53v42uvanduvbar} where 
the two polygons are obtained by reflection of one another \dots but for some extremal vertices that would fall
out of the dominant Weyl chamber. Three cases may occur:

\subsubsubsection{1. Case $\p\ge \r+\s$. }
In that case it follows from that inequality
and from the inequality $\x+\y\le \p+\q+\r+\s$ noticed above (end of sect.\,\ref{wesslen})
that $\x+\y\le 2\p+\q $; also from  
{the same inequality and from}
the second row of inequalities (\ref{8ineq}), it follows that 
$2\x+\y\le 2\p+\q+2\r+\s \le 5\p+4\q-\r-2\s$ and thus, in (\ref{transfo}), 
we take 
$\w'=\tr_1(\w)=(2\p+\q-\x-\y,\y)$.
It is then tedious but straightforward to check that if $\w$ satisfies the inequalities (\ref{8ineq}), then $\w'=\tr_1(\w)$
satisfies the inequalities (\ref{8ineq}) with $\v$ changed into $\bar\v$. Moreover as the transformation is involutive
and thus one-to-one,
it conserves the number of points: the  number of $\nu$ with $m=1$ (first layer of $\lambda\otimes\mu$)
equals that of $\nu'$ in the first layer of $\lambda\otimes\bar\mu$. See Fig. \ref{tens-poly-a} for an example.

\begin{figure}\begin{center}
\includegraphics[width=.5\textwidth]{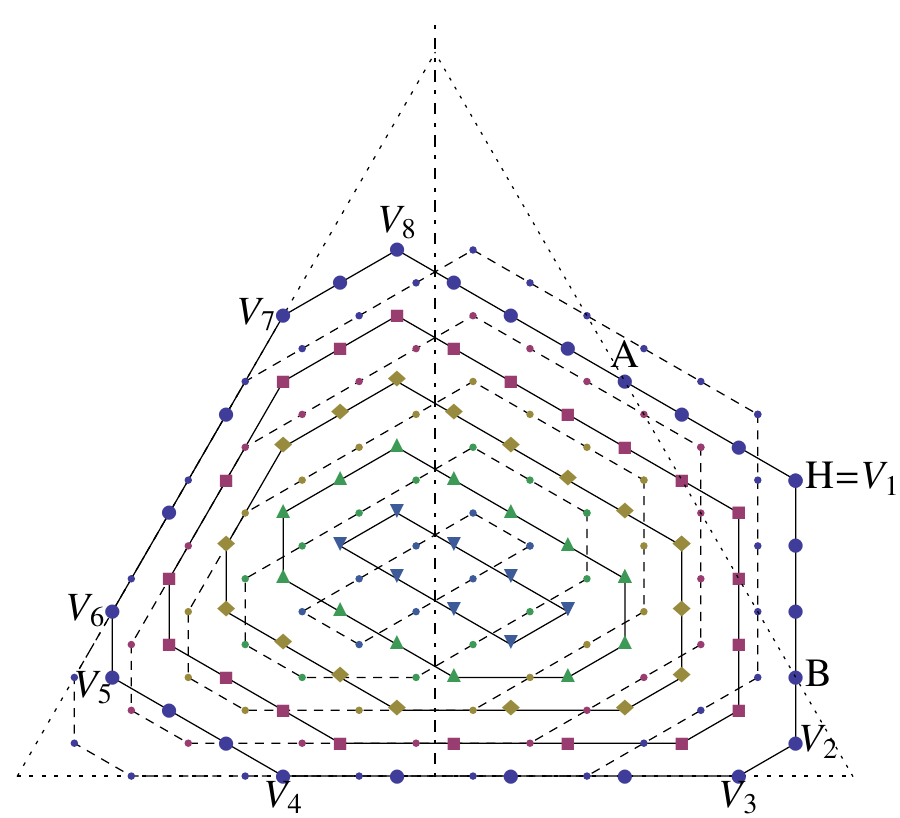}  
\end{center}
\caption{\footnotesize The two polygons for $(p,q,r,s)=(9,4,7,5)$ (solid) and $(9,4,5,7)$ (dashed), with the vertical reflection axis. }
 \label{poly9475} 
\end{figure}
\subsubsubsection{2. Case $\r+\s-\q\le \p< \r+\s$. }
In that case, one sees that one of the inequalities  (\ref{8ineq}) (in which $\mu$ has been changed into $\bar\mu$)
 may be violated, namely $\nu'_1 <0$ whenever $\nu_1+\nu_2> 2\lambda_1+\lambda_2$. The other inequalities are
 still satisfied. One thus modifies $\tr_1$, keeping a reflection in the domain $\nu_1+\nu_2\le 2\lambda_1+\lambda_2$,
 but changing it into a {\it translation} outside that domain
 \be\label{newt1} (\x',\y')=\tr_1'(\x,\y)=\begin{cases} \tr_1(\nu_1,\nu_2)=(2\p+\q-\x-\y,\y) & {\rm if\ } \x+\y\le 2\p+\q  \\ (\x,\y)+(\s-\r)(1,-1) & {\rm if\ } 
\x+\y> 2\p+\q\,. \end{cases}\ee
Again it is straightforward to verify that the resulting  $\nu'$ satisfy the right inequalities  (\ref{8ineq}), thus lie
inside or on the boundaries of  $P^{(1)}(\lambda,\bar\mu)$, and that the transformation is one-to-one. 
See Fig. \ref{poly9475} for an example, on which vertices   below the line AB are reflected by $\tr_1$, while those
above are shifted by $(-2,2)$.

\begin{figure}\begin{center}
\includegraphics[width=.4\textwidth]{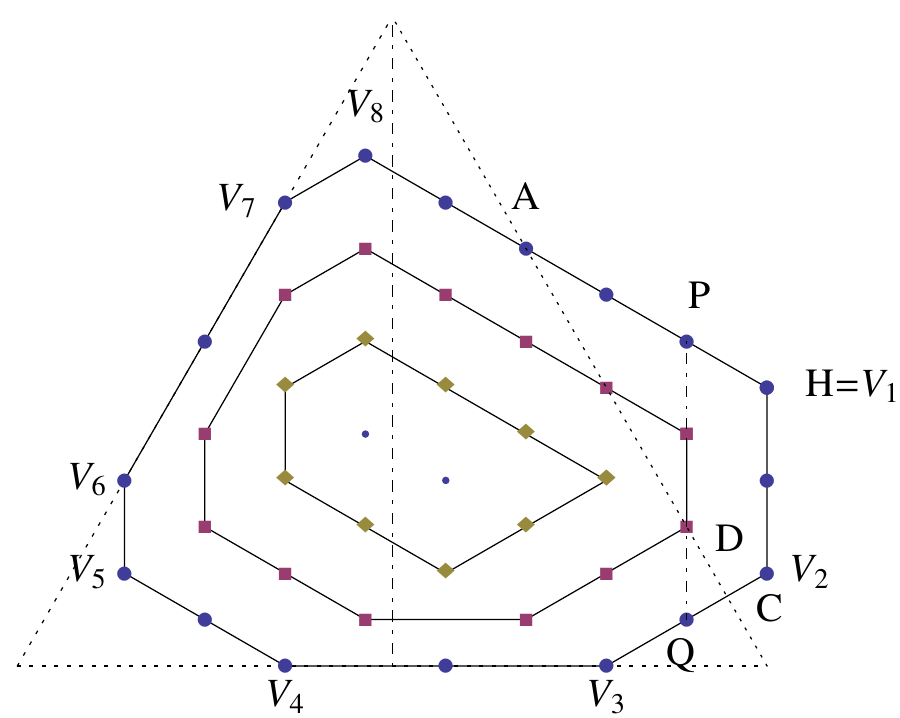}\qquad \includegraphics[width=.4\textwidth]{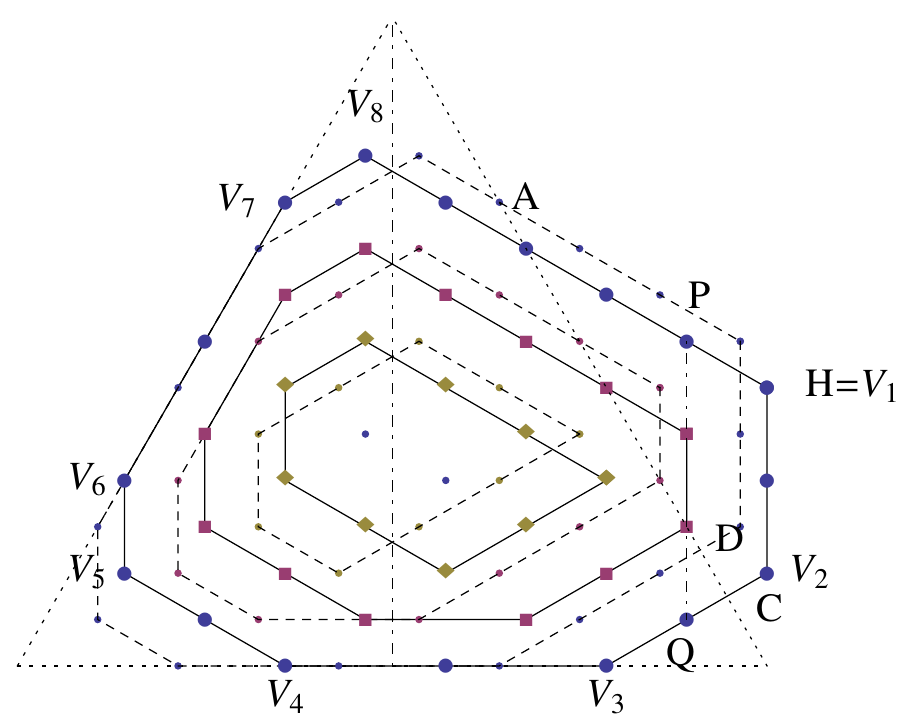}
\end{center}
\caption{\footnotesize The two polygons for $(p,q,r,s)=(6,2,5,4)$ (solid) and $(6,2,4,5)$ (dashed), with the vertical reflection axis. }
 \label{poly6254}
\end{figure}

\subsubsubsection{3. Case $\p< \r+\s-\q$.}
Finally in that case, another inequality, namely  (\ref{8ineq}), may be in trouble for $\nu'=\tr_1'(\nu)$:  $2\nu'_1+\nu'_2=3 \epsilon_1\cdot \nu' \ge 
\epsilon_3\cdot \lambda +\epsilon_1\cdot\bar\mu=\epsilon_3\cdot (\lambda -\mu) $ is not satisfied if
$\epsilon_1\cdot \nu> (2\epsilon_1-\epsilon_3)\cdot \lambda+ \epsilon_3\cdot \mu$ {\it i.e} $2\nu_1+\nu_2>
5\lambda_1+4\lambda_2-\mu_1-2\mu_2$. 
Thus for such $\nu$,  \ie  lying inside or on the boundaries of  $P^{(1)}(\lambda,\bar\mu)$ {\bf strictly above} the two lines
$\nu_1+\nu_2= 2\lambda_1+\lambda_2$ and $2\nu_1+\nu_2=
5\lambda_1+4\lambda_2-\mu_1-2\mu_2$, one modifies further  $\tr_1'$ into  $\tr_1''$, again by substituting 
a translation for the original reflection
\be\label{newt2}  (\x',\y')=\tr_1''(\x,\y)=\begin{cases} 
(\x,\y)+(\s-\r)(1,-1) & {\rm if\ }  2\x+\y> 5\p+4\q-\r-2\s\ \\ &{\rm or\ if\ } \x+\y> 2\p+\q  \\
\tr_1(\nu_1,\nu_2)=(2\p+\q-\x-\y,\y) & {\rm otherwise}\,.  \end{cases}\ee
There too, one checks that (\ref{8ineq}) is satisfied by $\nu'$  and that the transformation $\nu\mapsto \nu'$ is one-to-one. 
Formula (\ref{newt2}) is the case $m=1$ of (\ref{transfo}).
See Fig. \ref{poly6254} for an example: there, points below AD and left of PQ are reflected by $\tr_1$, while those
above AD or right of PQ are shifted by $(-1,1)$.

\subsubsection{The other layers}
To find the right transformation for the higher layers (values of $m>1$ in (\ref{mult-ind})), we make use 
of the relation (\ref{matriochK}) between the successive polygons and of the transformation (\ref{newt2}) on 
the first layer. This results in the formula (\ref{transfo}) and completes our discussion of the piecewise linear and 
layer-dependent  map of $\nu$ onto $\nu'$. 
\normalcolor

\section{Pictographs as combinatorial descriptions of the spaces of intertwiners}
\label{combinatorialmodels}

In this section we review various types of graphical representations of intertwiners, aka \pictograph s.
In the first subsection we gather several comments about the notion {of ``fundamental  \pictograph s''}.
There, we use the language of KT-honeycombs, defined in sect.\,\ref{proofusingKT},
 but the concepts could be equivalently described in terms of the other types of \pictograph s whose definition will be recalled or introduced in the second subsection.
In a third subsection, using the existence of fundamental \pictograph s,  
we shall obtain (still in the framework of SU(3)) a dictionnary between KT-honeycombs and these other \pictograph s.
In particular this dictionnary can be used to express the proof of section \ref{proofusingKT} in terms of the other models.

\subsection{Decomposition of \pictograph s } 
\label{DecompositionOfHoneycombs}
As we did not find any competing terminology in the literature to refer to the forthcoming concepts, we introduce the following terminology.\\
For $\SU(N)$, 
we call {\sl fundamental \pictograph s} the combinatorial models associated with intertwiners  of the type $f_1 \otimes f_2 \otimes f_3$ where  the $f_i$'s denote either a fundamental representation (such intertwiners may not exist) or the trivial one. The dimension of the corresponding spaces of intertwiners being equal to $0$ or $1$, there is only one such {\pictograph} or none.
 \\
For $\SU(N)$, we call {\sl non-primitive fundamental \pictograph s} the combinatorial models associated with the intertwiners  :
\begin{eqnarray*}
  f \otimes f^* \otimes \one&,&	 f ^*\otimes f \otimes \one, \\
  f \otimes \one \otimes f^*&,& 	 f ^*\otimes \one \otimes f,\\
  \one \otimes  f^* \otimes f&,&	\one  \otimes f \otimes f^*
\end{eqnarray*}  
where $f$ is a fundamental irrep. 
As the dimension of the corresponding spaces of intertwiners is equal to $1$, there is no ambiguity in this definition.
Clearly there are $3 \times (N-1)$ such \pictograph s, in particular six of them for SU(3).\\
For $\SU(N)$, we call {\sl primitive fundamental  \pictograph s} the combinatorial models associated with the intertwiners  :
$f_1 \otimes f_2 \otimes f_3$, where none of the $f_i$ is the trivial representation, and where the $f_i$ are fundamental representations.
It is easy to see that, if $G=\SU(N)$, there are $2 (N-2)(N-1)/2$ such \pictograph s, in particular two of them for SU(3), that are respectively associated with the cubes of $(1,0)$ and of $(0,1)$.

Typically, the external structure (``external legs'', in physicists parlance) of an object like a KT-honeycomb, \ie the set of numbers specifying the three chosen irreps, for instance the Dynkin labels of the corresponding highest weights, can be deduced from its internal structure, usually a set of numbers carried by the internal edges or by other graphical elements of the model. In the case of $\SU(N)$,  honeycombs or their like obviously contain $3 \times (1+2+\ldots+(N-1)) = 3 \times N(N-1)/2$ internal edges, in particular  $9$ edges for SU(3).\\
Arbitrary elements belonging to the vector space built over the set of internal edges cannot, in general, be considered as honeycombs, since the constraints that enter the definition of the latter would not, in general, be obeyed. Nevertheless, it is clear that one can add arbitrary honeycombs by adding the values of their internal edges, and multiply them by scalars.
\\
\begin{figure}[tbp]
\centering{\includegraphics[width=40pc]{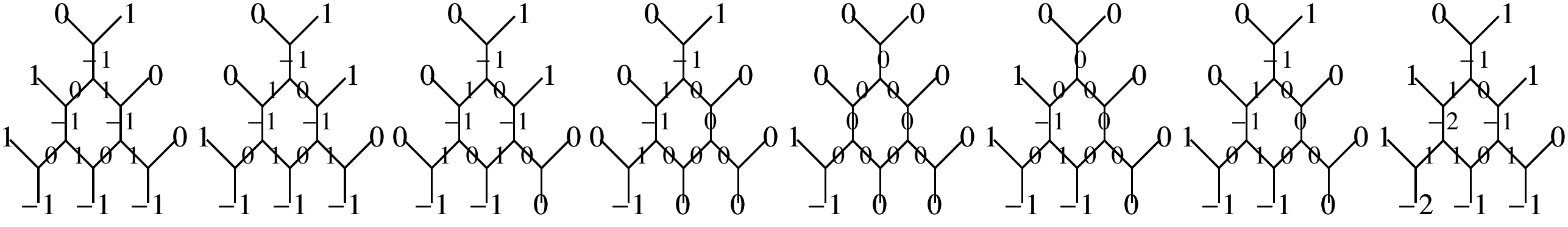}}
\caption{\label{fundamentalIntertwinersSU3a} Fundamental honeycombs 
for SU(3)}
\end{figure}

\begin{figure}[tbp]
\centering{\includegraphics[width=30pc]{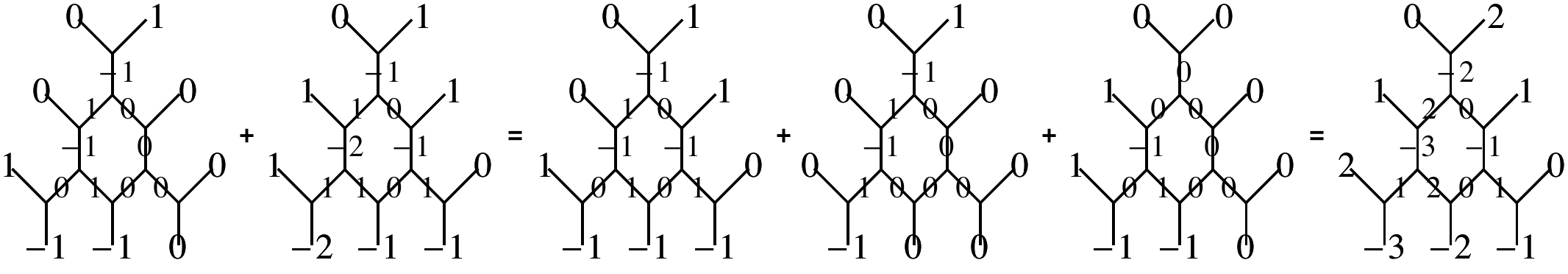}}
\caption{\label{su3RelationHoneycombs} The SU(3) intertwiner relation in terms of KT-honeycombs (left).
The second relation on the right is  one of the two honeycombs that describe the (2-dimensional) space of operators intertwining the square of the adjoint representation of SU(3)  with itself.}

\end{figure}

One could think that the previously described family of $(N-1)(N-2)+3(N-1)=(N^2-1)$ fundamental \pictograph s (primitive and non-primitive) can be chosen as a basis for the space of \pictograph s, but this is not so, as we shall see below, because they are not independent. In the case of SU(3) honeycombs for instance, we have $9$ internal edges, and $8$ fundamental honeycombs ($6$ non-primitive, $2$ primitive), see Fig. \ref{fundamentalIntertwinersSU3a},
 but there is also one relation
 which is displayed in Fig.  \ref{su3RelationHoneycombs} in terms of 
 KT-honeycombs.  The number of independent fundamental honeycombs is therefore $7$. 
Up to permutations, we can construct two particular basis, 
each  {one} contains seven fundamental intertwiners: the six non-primitive ones, and a last one chosen among the two that are primitive.

For $\SU(N)$ we have one relation for every inner hexagon (or for every inner vertex in the \oblade\ model, see below) \ie  $(N-2)(N-1)/2$ relations.
Formally, one may consider the vector space over the set of fundamental  
{\pictograph s}, a space of dimension $(N^2-1)$, and take its quotient by the relations. 
This quotient, the ``space of {\pictograph s}'',  has dimension $(N+4)(N-1)/2$.  Actually, in view of the correspondance between {\pictograph s} and {intertwiner spaces,}
  what matters are {\sl integral}Ê {\pictograph s}, but the fact that they are integral is usually understood. \\
Going back to our problem of relating the intertwiner spaces  $\u\otimes \v \to \w$  and $\u\otimes \overline{\v} \to \w^\prime$,  it is quite natural to consider the sets of associated {\pictograph s} 
and decompose them on a basis of (independent) fundamental ones. 
Notice  that the space of {\pictograph s}
comes with a distinguished generating family -- {those that are fundamental} -- 
but not with a distinguished basis as there is no canonical way to choose half of the  fundamental primitive {\pictograph s}
to build a basis.

\subsection{Other {\pictograph s}: 
BZ-triangles, \oblade s or SU(3)-honeycombs}
\subsubsection{BZ-triangles}

\begin{figure}[ht]
\centering{\includegraphics[width=12pc]{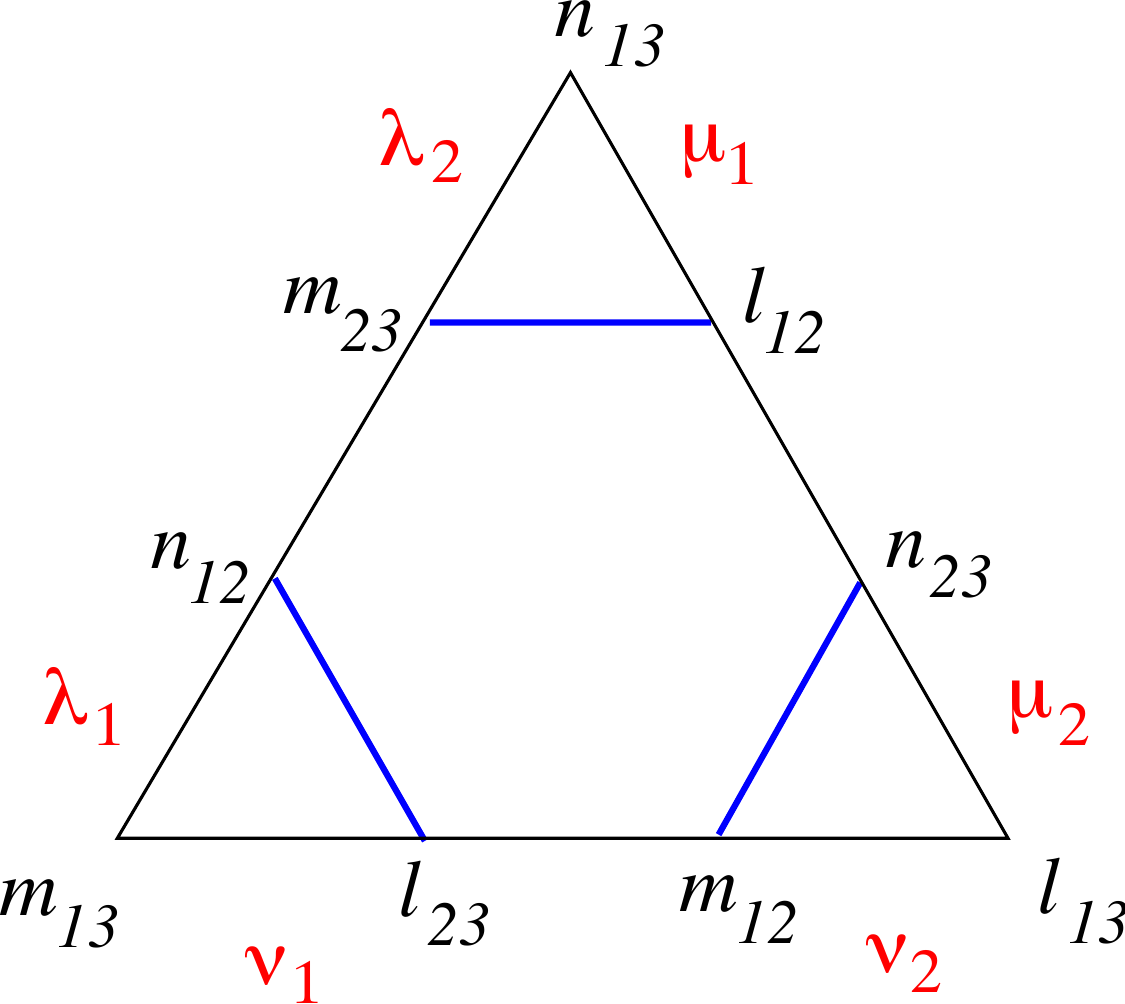}}
\hspace{3pc}
\centering{\includegraphics[width=14pc]{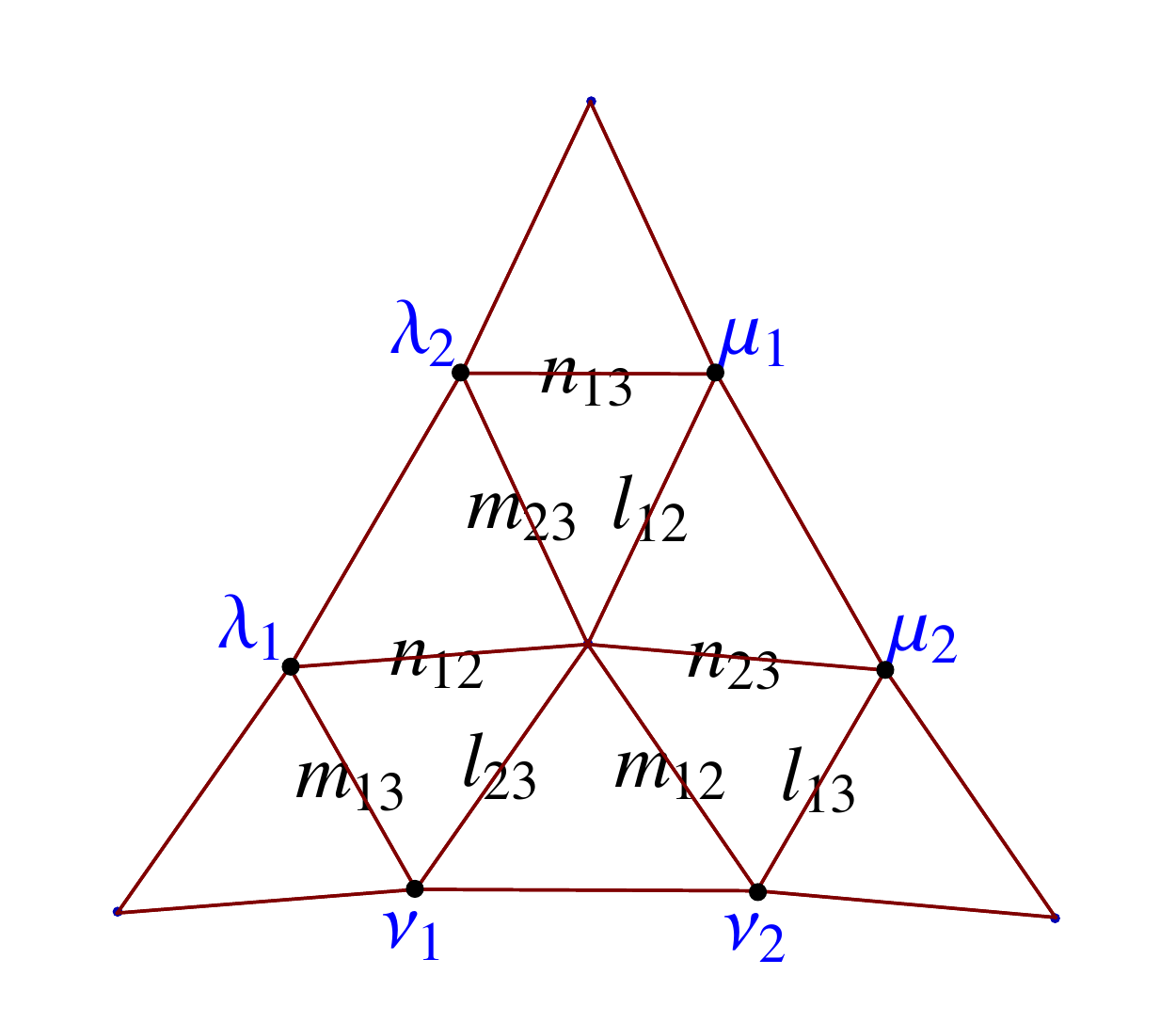}}\\
{{(a)}\hskip7cm{(b)} }
\caption{\label{genericBZtriangleAnsObladeSU3}Generic BZ-triangle 
and \oblade\  for SU(3)}
\end{figure}

As Berenstein--Zelevinsky (BZ)-triangles are well documented in other places \cite{BZ}, 
we shall be happy with the following operational definition, adapted to the case of SU(3).
A generic BZ triangle for SU(3) is displayed in Fig.\,\ref{genericBZtriangleAnsObladeSU3} (a). It  contains nine vertices  called
$m_{12}, m_{23}, m_{13}, n_{12}, n_{23}, n_{13}, l_{12}, l_{23}, l_{13}$
labelled with non-negative integers.   
The BZ-triangle for SU(3) has a single inner hexagon, and  obeys one single constraint:
 the opposite sides of the hexagon should be equal (the value of a side being defined as the sum of its endpoints).
 This reads: 
\begin{eqnarray}\label{hex-constr}
 l_{12} + n_{23} &=& l_{23} + n_{12} \cr
 m_{12} + l_{23} & = &m_{23} + l_{12} \\
\nonumber n_{12}+ m_{23} &=& n_{23} + m_{12} \,.
\end{eqnarray}
Given a BZ-triangle, \ie a  set of nine non-negative integers obeying the above constraint, one defines the following six non-negative integers: 
\begin{eqnarray*}
\lambda_1 &=& m_{13} + n_{12}  \qquad  \mu_1 = n_{13} + l_{12}  \qquad   \nu_1 = m_{13} + l_{23}\\
\lambda_2 &=& m_{23} + n_{13}  \qquad  \mu_2 = n_{23} + l_{13}   \qquad  \nu_2 = m_{12} + l_{13} \,.
\end{eqnarray*}
In the case of SU(3), the BZ theorem can be expressed as follows:  the dimension of the space of intertwiners associated with the branching rule $\lambda \otimes \mu \mapsto \nu$ (or, equivalently $\lambda \otimes \mu \otimes \overline{\nu} \mapsto 1$, or equivalently,  the multiplicity of the irrep $\nu$ in the tensor product $\lambda \otimes \mu$) is equal to the number of BZ-triangles obeying the above set of constraints. As this is discussed elsewhere, let us just mention that the nine integers entering a BZ-triangles are the components of the (non-unique) decomposition of the three vectors\footnote{The writing of the three $\sigma_i$ looks more symmetrical if one uses $\overline{\nu}$ instead of $\nu$ in the definition.} $\sigma_1 = \lambda+\mu-\nu$,  
$\sigma_2 = \lambda - \overline{\mu}+\overline{\nu}$, $\sigma_3=-\overline{\lambda} + \mu+\overline{\nu}$,  belonging to the SU(3) root lattice,  along the generating family (not a basis) of positive roots (a triplet in this case).  More generally, for $\SU(N)$, we have $N(N-1)/2$ positive roots, hence $3 \times N(N-1)/2$ vertices in the BZ-triangles,  each inner hexagon determining the constraints that are obeyed by the integers associated with vertices.

\begin{figure}[tbp]
\centering{\includegraphics[width=38pc]{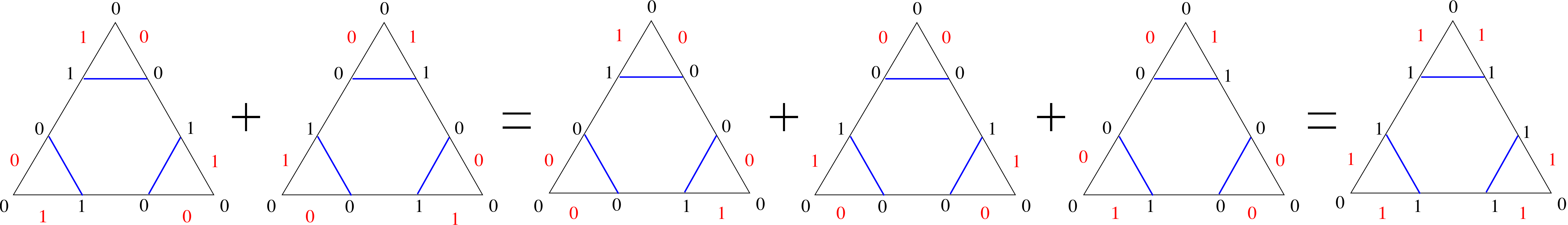}} 
\caption{\label{relationforBZ} The SU(3)  intertwiner relation, in terms of BZ-triangles }
\end{figure}

\begin{figure}[htbp]
 \centering{
  \includegraphics[width=10pc]{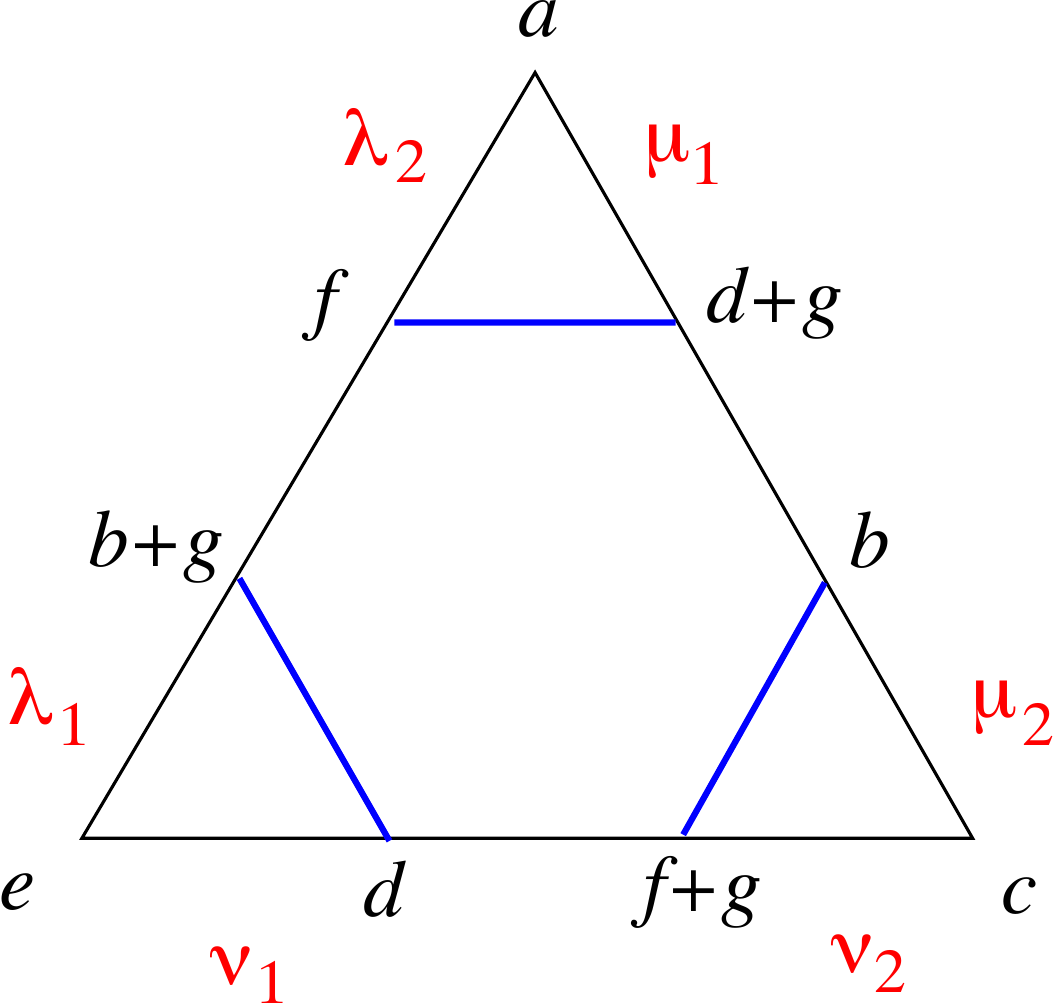}\qquad
 \includegraphics[width=12pc]{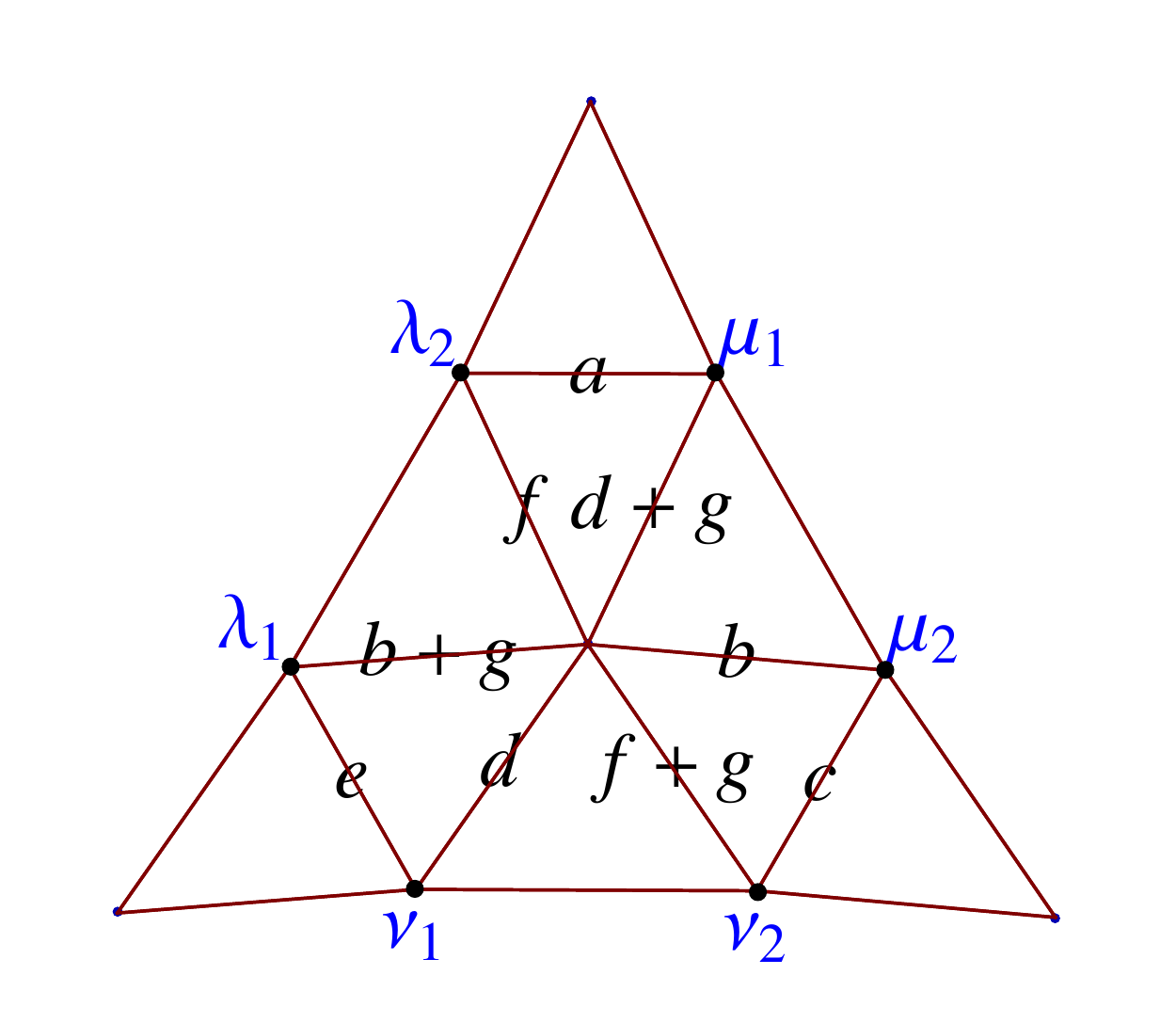}}\\
 {(a)}\hskip50mm{(b)} 
\caption{\label{arbitraryobladeortriangleSU3} The seven components $(a,b,c,d,e,f;g)$ of a BZ-triangle, or of an \oblade\  for SU(3)}
\end{figure}

 Thought of as $9$-plets, SU(3) BZ-triangles can be linearly combined, or multiplied by scalars; however, and as it was discussed in 
 sect. \ref{DecompositionOfHoneycombs}, any BZ-triangle can be written -- in a non-unique way -- as a linear combination of $8$ {\sl fundamental} BZ-triangles. The non-uniqueness comes from the fact that, for SU(3), there is one linear relation between the fundamental triangles. This relation, displayed\footnote{It can be expressed in terms of the other combinatorial models (see section \ref{DecompositionOfHoneycombs} and Fig. \ref{su3RelationHoneycombs} and \ref{su3RelationOblades}).} in Fig.  
 \ref{relationforBZ}, can be immediately checked. 
 One can then select, for instance,  a basis (see our discussion in section \ref{DecompositionOfHoneycombs}) 
 made of  the six non-primitive fundamental triangles, and a last one,  chosen among the two that are primitive.
  It is therefore enough to choose $7$ arbitrary non-negative integers\footnote{Only the {\sl integral} BZ-triangles have an interpretation in terms of intertwiners.}  to build arbitrary BZ-triangles obeying all the constraints.  
 If $(a,b,c,d,e,f;g)$ denote those seven  components, one sees immediately that an arbitrary BZ-triangle
  is given by Fig.  \ref{arbitraryobladeortriangleSU3} (a), and that, with this parametrization, the hexagon constraint 
  (\ref{hex-constr}) becomes automatic.
  
\subsubsection{\oblade s (or oblades)}

The name ``oblade'' that we  use in the following to denote an alternative kind of combinatorial model is not supposed to be understood as a fish ({\it  Oblada melanura}, a seabream) but as an \oblade, because, to the best of our knowledge, this way of displaying the BZ equations is due to A. Ocneanu \cite{AOblades},  whom we want to thank for having shared this viewpoint with us.

\begin{figure}[tbp]
\centering{\includegraphics[width=42pc]{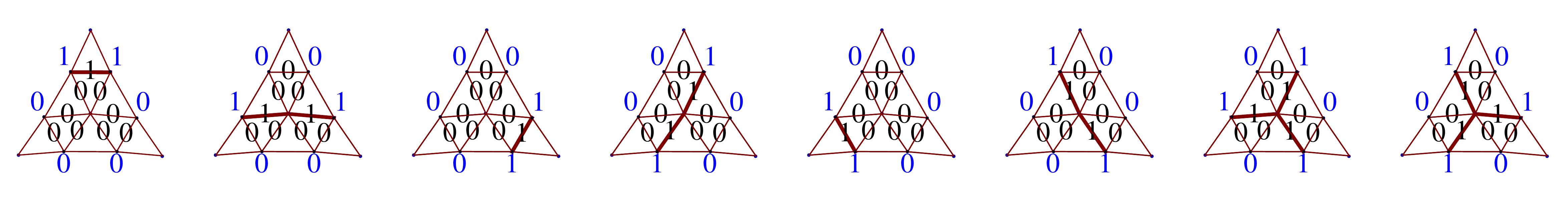}}
\caption{\label{fundamentalIntertwinersSU3} Fundamental \oblade s for SU(3)}
\end{figure}

\begin{figure}[tbp]
\centering{\includegraphics[width=35pc]{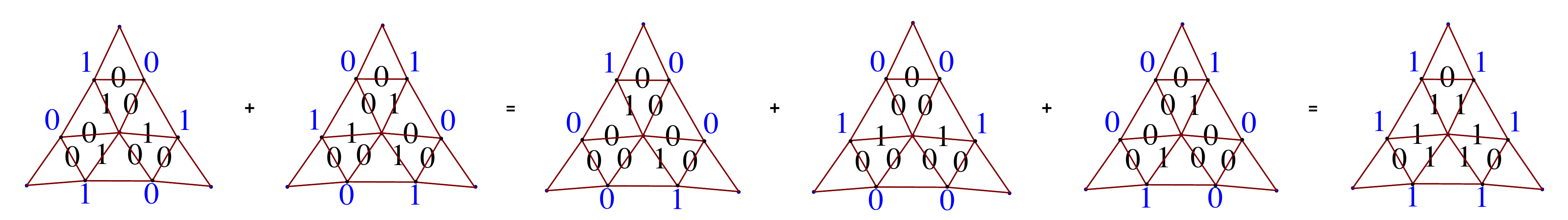}}
\caption{\label{su3RelationOblades} The SU(3) intertwiner relation in terms of \oblade s }
\end{figure}

\begin{figure}[tbp]
\centering{\includegraphics[width=35pc]{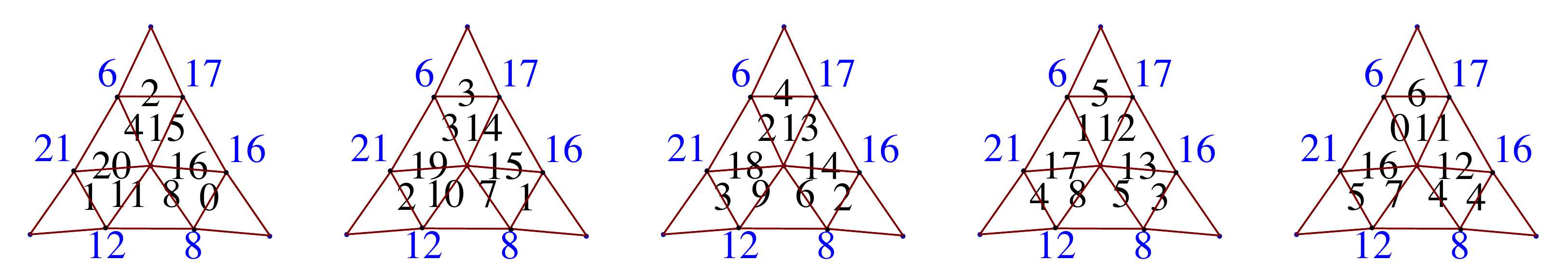}}
\caption{\label{oblades2161716128} The five \oblade s of the case $(21,6) \otimes (17,16) \rightarrow (12,8)$}
\end{figure}

A generic \oblade\  for SU(3) is displayed in Fig. \ref{genericBZtriangleAnsObladeSU3} (b). \\ 
It  contains nine edges
$m_{12}, m_{23}, m_{13}, n_{12}, n_{23}, n_{13}, l_{12}, l_{23}, l_{13}$,
 labelled with non-negative integers.  \\
 There is an obvious geometrical way to establish a one-to-one correspondence between this combinatorial model and the previous one:
{the BZ triangle of Fig.\,\ref{genericBZtriangleAnsObladeSU3} (a) and its labels  
  are read off the inner edges of the \oblade\  of  Fig.\,\ref{genericBZtriangleAnsObladeSU3} (b).}
\\
Because of this correspondence, everything that was written in the previous section can be immediately translated in this other language.
For instance, the \oblade\  for SU(3) has a single inner vertex, and  obeys one single constraint:
the three pairs of opposite ``angles" defined by the lines intersecting at the inner vertex should be equal, the value of an angle being defined as the sum of its sides.
The equations themselves read the same as in the previous case. As for BZ-triangles, the three weights $\lambda, \mu, \overline{\nu}$ follow each other cyclically,  in a clockwise manner (this is our choice), so that the components of the weights appearing in the branching rule $\lambda \otimes \mu \rightarrow \nu$ are as indicated on the pictures.
 The eight fundamental \oblade s are given in Fig.  \ref{fundamentalIntertwinersSU3} {-- edges carrying a ``$1$'' label have been thickened --}  with the corresponding KT-honeycombs already given in Fig.\ref{fundamentalIntertwinersSU3a} (the interested reader will have no difficulty in drawing the corresponding BZ-triangles). Notice that the two primitive fundamental \oblade s appear as Y-shapes\footnote{The reader may certainly devise a tropical interpretation for honeycombs and their like\ldots} (forks) with a horizontal tail to the left or to the right of the  inner vertex. The unique relation between them is displayed  in Fig.   \ref{su3RelationOblades}. 
Finally,  chosing again the ``left'' basis (\ie the last basis element being Y-shaped,  with a horizontal tail to the left of the inner vertex), 
we see in Fig.\,\ref{arbitraryobladeortriangleSU3} (b) how an arbitrary \oblade\  can be defined as a superposition of its seven components.
In terms of \oblade s, the five KT-honeycombs discussed at the beginning of sect.\,\ref{proofusingKT} and that are associated with the space of intertwiners specified by the  branching rule $(21,6)\otimes(17,16) \rightarrow (12,8)$, read as in Fig. \ref{oblades2161716128}. 
   
\subsubsection{SU(3)-honeycombs}

\begin{figure}[tbp]
\centering{\includegraphics[width=15pc]{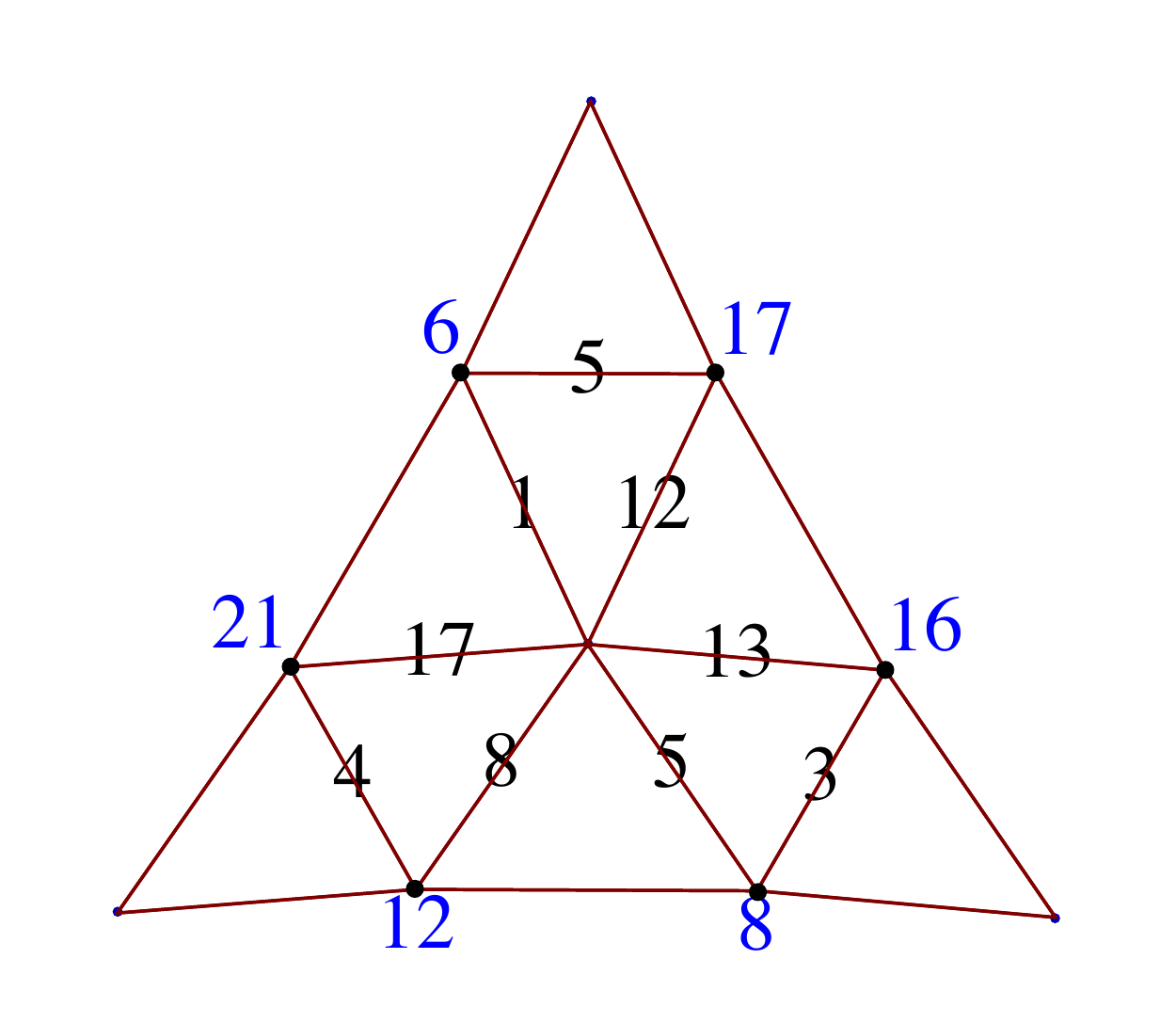}}\qquad
\centering{\includegraphics[width=13pc]{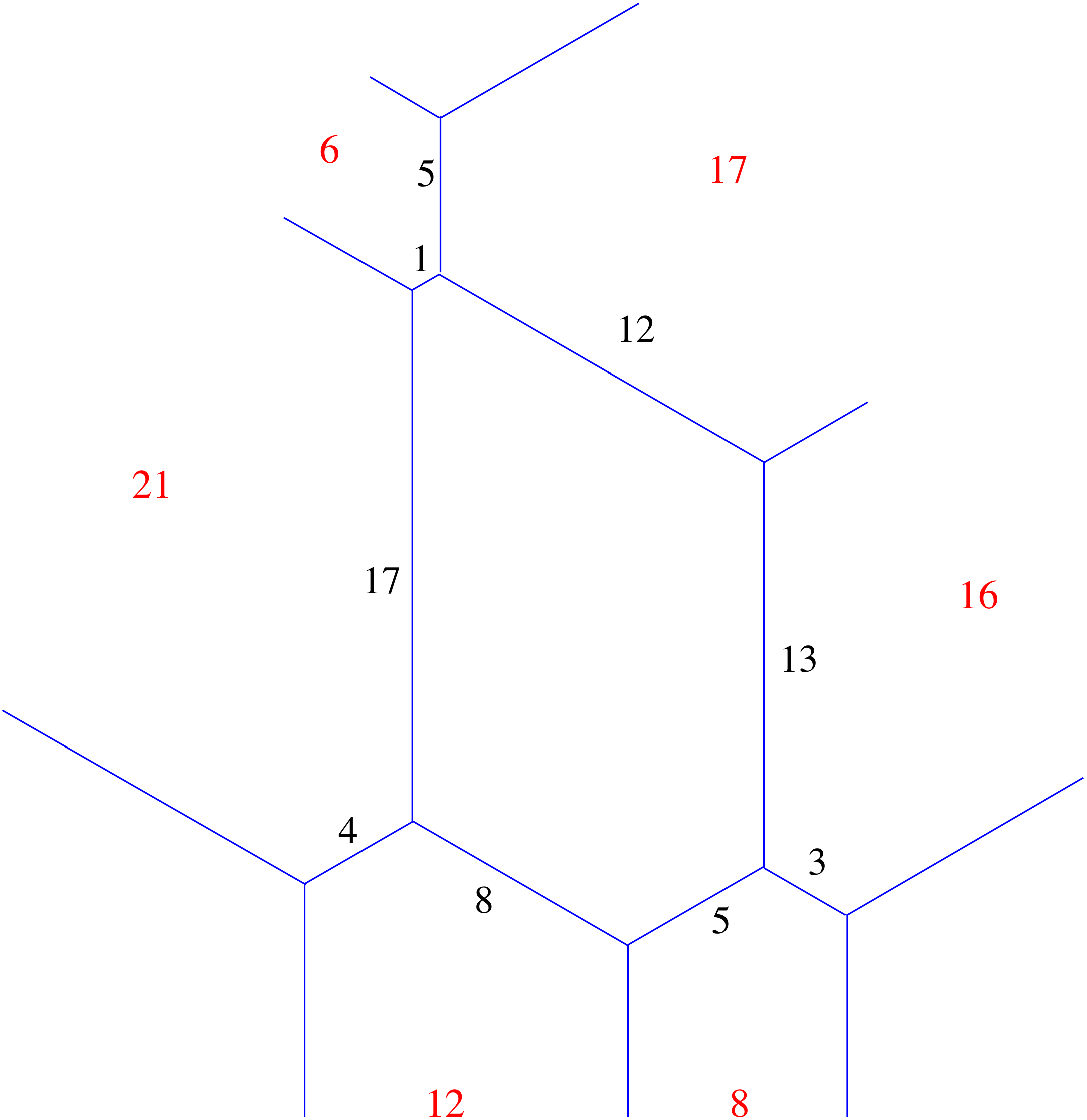}}
\caption{\label{ObladesAndSL3Honeycombs} An \oblade\  with its (dual) SU(3)-honeycomb drawn in a metric way
}
\end{figure}

The dual of an \oblade\  is obviously a honeycomb. The geometrical relation between the two should be clear from Fig.  \ref{ObladesAndSL3Honeycombs} which displays one of the corresponding pairs associated with the  intertwiners
defined by the already discussed branching rule $(21,6)\otimes(17,16) \rightarrow (12,8)$.
These ``SU(3)-honeycombs'' are not ``KT-honeycombs" 
since the integers labels, and the constraints they obey, are different. In particular the lengths of all edges are non-negative. Also,  the external labels refer to Dynkin labels of SU(3) highest weights, and are therefore also non-negative,  whereas in the case of KT-honeycombs, the external indices refer to partitions labeling GL(3) irreps.
The translation of BZ-triangles, or of \oblade s, to this last combinatorial model is immediate, and we leave this  as an exercise for the reader. 
As a side remark, note that one of the merits of these {SU}(3)-honeycombs is to be ``metric", in the sense that with angles of 
$2\pi/3$, their edges may be drawn with the indicated value, see Fig.  \ref{ObladesAndSL3Honeycombs}.

The purpose of this short section was only to avoid possible misunderstandings and warn the reader that one can devise several types of honeycomb \pictograph s. 
 In the present paper all calculations involving combinatorial models were done using KT-honeycombs or \oblade s, but ``{\it De gustibus non est disputandum}''.

  \subsubsection{Virtual pictographs}
  
  All the pictographs used or mentioned in this paper obey the positivity constraints already mentioned.   
  However, it may be sometimes useful to introduce pictographs that are not of that kind.
  For instance, in the case of SU(3), one can introduce a ``virtual O-blade'' $\Delta$, with edges $m_{ij}, n_{ij},  \ell_{ij}Ê$ 
  equal to $(-1,-1,1, -1,-1,1, -1, -1, 1)$,  with the notation of section 4.2.2, 
  and as in Fig.  \ref{genericBZtriangleAnsObladeSU3}.
  The corresponding representations (external labels) are $(0,0), (0,0), (0,0)$, i.e., three times the trivial representation.
  In terms of BZ-triangles, this virtual pictograph was shown in \cite{BMW} to play the role of a step-operator between the various pictographs associated with the same branching rule  (see also a similar discussion in \cite{DFMS}).
  Indeed, all the pictographs $\Delta_n$, $n = 0, 1, \ldots, \mult(\lambda, \mu; \nu)-1$  associated with a given triple 
  $(\lambda, \mu, \nu)$ of highest weights are of the kind $\Delta_n = \Delta_0 + s  \Delta $, 
  where $s=1,2,\ldots \mult(\lambda, \mu; \nu)$.


\subsection{From BZ-triangles, \oblade s and SU(3)-honeycombs to KT-honeycombs and hives: Yet  
another approach to the permutation property}

As it was discussed in the previous section, the relation between the graphical components belonging to the first three combinatorial models appearing in the title of this section is immediate.
Although slightly less obvious, the relation between the latter and KT-honeycombs or hives is not difficult
to grasp. One way to obtain this relation is to decompose an arbitrary \oblade\  (for instance) and an arbitrary KT-honeycomb on a basis of fundamental ones, and compare the results. Indeed the matching  between 
 fundamental KT-honeycombs and \oblade s is one-to-one,
  and it is shown explicitly in the corresponding figures \ref{fundamentalIntertwinersSU3a} and \ref{fundamentalIntertwinersSU3}. 
As before, let us assume that we have chosen some
basis which contains the six non-primitive  ones and one of the two that are primitive, and 
call $(a,b,c,d,e,f;g)$ the components along that basis.
The values of the nine edges of the \oblade\  (or of the nine vertices of the BZ-triangle), in terms of $(a,b,c,d,e,f;g)$ are  immediately obtained. The result was already given (compare Fig.  \ref{genericBZtriangleAnsObladeSU3}~(b) and \ref{arbitraryobladeortriangleSU3} (b). The nine edges of the KT-honeycomb (read from top to bottom, and left to right, see Fig.  \ref{KThoneycombSL3}), in terms of these seven parameters,  read respectively:
$$(-a - b - c - d - g, b + c + d + g, a, -a - b - c - d - f - g, -a - b - c, c + d, a + b + f + g, c, a + b)\,.$$
In terms of the same components, the three external GL(3) weights (partitions) read:  $$\{e + b + g + a + f, a + f, 0\}, \{a + d + g + b + c, b + c, 0\}, \{-(e + d + f + g + c) - (a + b), -(f + g + c) - (a +  b), -(a + b)\}\,.$$
Finally, the ``hive parameter'' called $\a$ 
 everywhere in sect.\,\ref{proofusingKT} (also in the figures of that section), is simply related to first parameter $a$, of the list $(a,b,c,d,e,f,g)$, one finds:    $\lambda_1+2 \lambda_2 + \mu_1+\mu_2  - \a  =  a$.  For instance, in the case of the example 
 we found that $\a$ runs from $60$ to $64$, so that the corresponding $a=66-\a$ runs  from $2$ to $6$.
Notice that $a = n_{13}$ is the edge located at the top of the \oblade\  (or at the top vertex of the BZ-triangle).

Using this dictionnary, the reader will have no difficulty to re-express the $9$ ``Wesslen inequalities'' of section \ref{wesslen},   already interpreted in terms of KT-honeycombs, 
as positivity constraints for the $9$ edges (``blades'') of the \oblade s, and, more generally, to  recast the whole analysis carried out in the two previous sections in terms  of the other combinatorial models.

\section{Final comments.}
\label{final-comments}

We have observed that theorem \ref{CZthm2}, proved for SU(3), also holds for the fusion of representations of the affine algebra $\widehat{\su}(3)_k$ at a finite level $k$.
This, however, would require a separate proof that we actually did not complete, and that in any case would markedly increase the size of the paper.
By way of contrast, the same properties do not extend, in general, to higher rank algebras or groups
endowed with complex representations.  We gather a few comments on the case SU(4).

For the affine algebra $\widehat{\su}(4)_k$, theorem \ref{CZthm2}
holds true for all weights of levels less or equal  to 7, but fails for some weights at levels greater or equal to 8. 
For instance, both for $\widehat{\su}(4)_8$  and for the classical group SU(4),  a weight of multiplicity 8 appears in  the decomposition of the tensor product $(1,2,2)\otimes (2,1,3)$,  
while all multiplicities are bounded by 7 in the decomposition of $(1,2,2)\otimes (3,1,2)$. On the other hand, 
theorem \ref{CZthm2} is still satisfied by many pairs $(\lambda,\mu)$, but we have been 
 unable to find a simple criterion for this to hold. 
It is nevertheless interesting to ask: what goes wrong in the previous proofs, when one passes from SU(3) to SU(4) ?

In the case of $\SU(N)$,  the pictographs obtained in sect. 4 for given external $\lambda$, $\mu$ and $\nu$ depend on $(N-1)(N-2)/2$ 
parameters rather than on a single one, $\a$, as in the case of SU(3). As a result, the determination of possible values of 
these parameters, \ie the multiplicity, becomes more delicate since it typically amounts to count
integral points in a convex set, a notoriously difficult problem. The analysis that we could perform for SU(3) cannot therefore be extended in a straightforward manner and such a generalization would not give rise to the same conclusions. 

\begin{figure}[tbp]
\centering{\includegraphics[width=15pc]{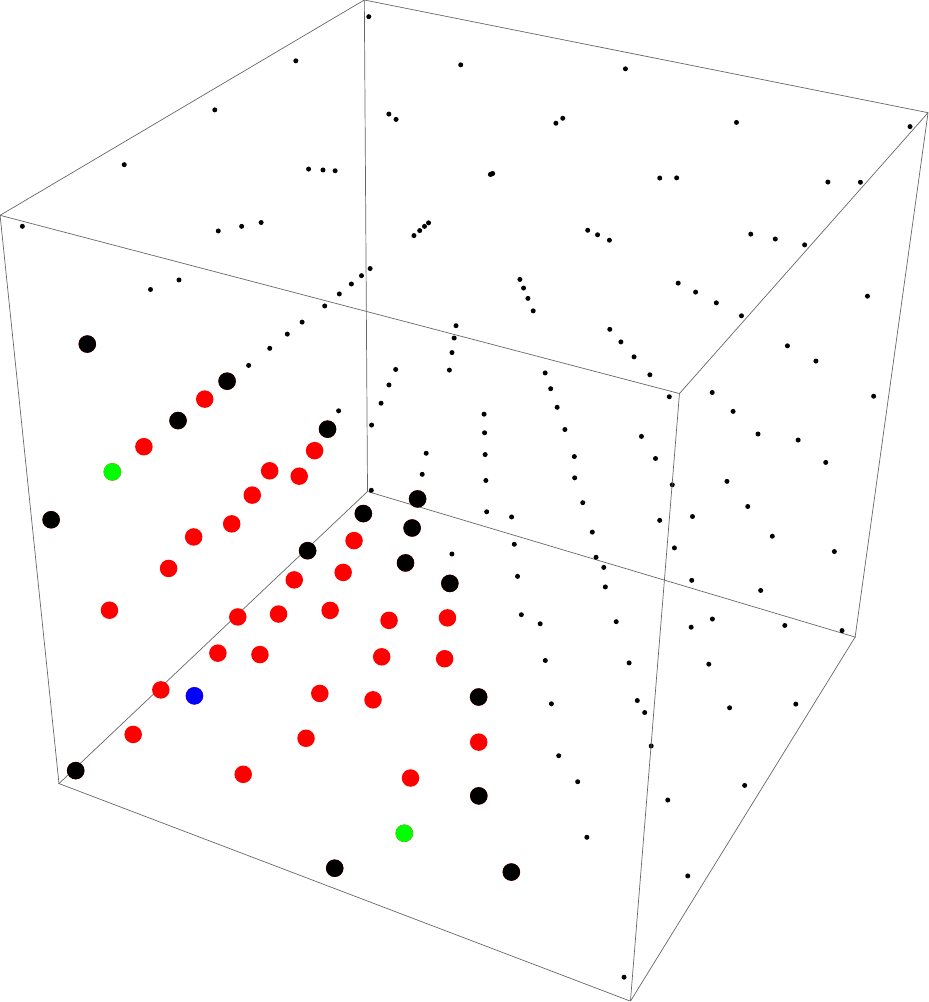}}
\caption{\label{polytopesu4} 
 An SU(4) example: decomposition of $\lambda \otimes \mu$ with $\lambda=(1, 2, 2)$ and $\mu= (2, 2, 1)$.
Axes are along the $\omega_1,\omega_2,\omega_3$ 
directions.  The black dots represent weights of multiplicity 1 and are the vertices of a 
convex polytope.  The  two weights of multiplicity 2,  in green,
and  the weight of multiplicity 5, in blue, lie in the plane $\nu_2=0$, one of the faces of the polytope. 
The other weights of the decomposition are in red. Thin dots are points of the shifted root lattice
$\lambda+\mu+Q$.}
\end{figure}

It is also interesting to observe that the  convexity property of domains, like the ``layers" that appeared in SU(3),
 is less stringent for SU(4) and beyond : the polytopes of increasing multiplicity may touch one another. An instance is presented in Fig. \ref{polytopesu4}:
the weights $\nu$ appearing  in the decomposition of $(1, 2, 2)\otimes (2, 2, 1)$ form a  polytope determined by the (black) vertices \\
{$\scriptstyle \{(0, 0, 0), (0, 0, 4), (4, 0, 0),  (6, 1, 0),  (0, 1, 6),  (0, 6, 0), 
(2, 0, 6), (6, 0, 2), (4, 4, 0),(0, 4, 4), 
(5, 0, 5), (3, 4, 3), (1,5,3), (3,5,1), (1,6,1), (4,2,4)\}$}, \\ all of multiplicity 1; 
however, on the face determined by vertices 
{$\scriptstyle (0, 0, 0), (0, 0, 4), (2,0,6), (6,0,2),(4, 0, 0)$}, there are weights of higher multiplicity, e.g. {$\scriptstyle (5,0,1), (1,0,5)$} which are of multiplicity 2 (green), and {$\scriptstyle (2,0,2)$} of multiplicity 5 (blue).

\section*{Acknowledgements} 

It is a pleasure to thank Allan Knutson and Oleg Ogievetsky for stimulating discussions. Partial support from ESF ITGP network is acknowledged.


\vfill \eject


\begin{thebibliography}{99}
\footnotesize

\bibitem{CZ1} R. Coquereaux and  J.-B. Zuber, {\sl On sums of tensor and fusion multiplicities}, J. Phys. A {\bf 44} (2011) 295208.

\bibitem{Mandeltsveig} V.B. Mandel'tsveig, {\sl Irreducible representations of the $SU_3$ group}, Soviet Physics JETP 
{\bf 20} (1965), 1237--1243. 

\bibitem{Wesslen} M. Wesslen, {\sl A geometric description of tensor product decompositions in su(3)},  
J. Math. Phys. {\bf 49} (2008) 073506.


  \bibitem{DFMS} P. Di Francesco, P. Mathieu and D. S\'en\'echal, {\sl Conformal Field Theory}, Springer 1997.

   
  \bibitem{PRV}  K.R. Parthasarathy, R. Ranga Rao, V.S. Varadarajan,  {\sl Representations of complex semi-simple Lie groups and Lie
algebras}, Ann. of Math. (2) {\bf  85} (1967), 383--429.

\bibitem{Kumar} S. Kumar,  
 {\sl A refinement of the PRV conjecture},
 Invent. Math {\bf 93} (1988), 117--130.

\bibitem{Mathieu} O. Mathieu, 
 {\sl Construction d'un groupe de Kac-Moody et applications},  
Composition Mathematica {\bf 69} no1  (1989), 37--60.
   
   \bibitem{MPR}  PL. Montagard, B. Pasquier, N. Ressayre 
   {\sl Two generalizations of the PRV conjecture},
  Compositio Math. {\bf  147} (2011), 1321--1336;
   arXiv:1004.2119v1.
   
    \bibitem{FrenkelZhu} I. B. Frenkel, Y. Zhu, 
     {\sl Vertex operator algebras associated to representations of affine and Virasoro algebras}, 
    Duke Math. J. {\bf 66}  (1992), 123--168.
    
\bibitem{FeingoldFredenhagen} 
A. J. Feingold and S. Fredenhagen,
 {\sl A New Perspective on the Frenkel-Zhu Fusion Rule Theorem}, 
Journal of Algebra {\bf 320} (2008) 2079--2100;  arXiv:0710.1620v1 [math.RT].


\bibitem{skYoung}  W. Fulton,  {\sl Young Tableaux, with Applications to Representation Theory and Geometry},  Cambridge University Press 1997. 

\bibitem{BZ} A. Berenstein and A. Zelevinsky, {\sl Triple multiplicities for sl($r+1$) and the spectrum of the external algebra in
the adjoint representation}, J. Alg. Comb. {\bf 1} (1992), 7--22.

\bibitem{KT1} A. Knutson and T. Tao, {\sl The honeycomb model of $\mathrm{GL}_n(\mathbb{C})$ tensor products I: 
Proof of the saturation conjecture}, 
J. of the AMS {\bf 12} (1999) 1055--1090; \\ A. Knutson, T. Tao and C. Woodward,  {\sl The honeycomb model of $\mathrm{GL}_n(\mathbb{C})$ tensor products II: Puzzles determine facets of the Littlewood--Richardson cone}, J. of the AMS
 {\bf 17} (2003), 19--48;\\
 A. Knutson and T. Tao, {\sl Honeycombs and sums of Hermitian matrices}, 
 Notices Amer. Math. Soc. {\bf 48} (2001), no. 2, 175Ð186.
 arXiv:math/0009048 [math.RT]. 

\bibitem{KT2} 
A. Knutson and T. Tao, {\sl Puzzles and (equivariant) cohomology of Grassmannians}, Duke Math. J. {\bf 119} no. 2  (2003),  221--260. arXiv:math/0112150 [math.AT].

\bibitem{AOblades} A. Ocneanu (private communication, 2009).

 \bibitem{Littelmann}  {P. Littelmann}, {\sl A Littlewood--Richardson rule for symmetrizable Kac--Moody algebras}, 
 Invent. math. {\bf 116} (1994), 329--346;  
{\sl Paths and root operators in representation theory}, 
 Ann.  Math.  {\bf 142} (1995), 499--525. 
 
\bibitem{BMW}  { L. B\'egin, P. Mathieu and M.A. Walton},  
   {\sl $\widehat{su}(3)_k$ fusion coefficients},
  Mod. Phys. Lett. A  {\bf 7} No. 35 (1992);
  hep-th/9206032.
  
\bibitem{Suciu} L. Suciu. \textit{The SU(3) wire model}, PhD thesis,  The Pennsylvania State University, 1997; AAT 9802757.




\end{thebibliography}
 \end{document}